\documentclass[final,5p,times,twocolumn]{elsarticle}
\usepackage{graphics}
\usepackage{amssymb}
\usepackage{url}
\usepackage{graphicx}
\usepackage{soul}
\usepackage{amsthm}
\usepackage{multicol}
\usepackage{textcomp}
\usepackage{booktabs}
\usepackage{amsmath}
\usepackage[draft, inline, nomargin]{fixme}
\fxsetup{theme=color, mode=multiuser}
\FXRegisterAuthor{ab}{1}{\color{blue}AB}
\usepackage{lineno}
\usepackage{threeparttable}
\usepackage[colorlinks]{hyperref}
\usepackage{overpic}

\biboptions{sort&compress}

\journal{Nuclear Instruments and Methods in Physics Research, Part A}

\newcommand{\deflen}[2]{%
    \expandafter\newlength\csname #1\endcsname
    \expandafter\setlength\csname #1\endcsname{#2}%
}

\begin{document}

\begin{frontmatter}

\title{Evaluation of a Positron-Emission-Tomography-based SiPM readout\\for Compact Segmented Neutron Imagers}

\author[LLNL]   {V.~A.~Li}\ead{li68@llnl.gov}
\author[LLNL]   {F.~Sutanto}\ead{sutanto2@llnl.gov}
\author[LLNL]   {T.~M.~Classen}
\author[LLNL]   {S.~A.~Dazeley}%
\author[UM]     {I.~Jovanovic} 
\author[UM]     {T.~C.~Wu}
\address[LLNL]{Lawrence Livermore National Laboratory, Livermore, CA 94550}
\address[UM]{Department of Nuclear Engineering and Radiological Sciences, University of Michigan, Ann Arbor, MI 48109}

\begin{abstract}

Gamma-ray emission from special nuclear material (SNM) is relatively easy to shield from detection using modest amounts of high-Z material. In contrast, fast-neutrons are much more penetrating and can escape relatively thick high-Z shielding without losing significant energy. Furthermore, fast neutrons provide a clear and unambiguous signature of the presence of SNM with few competing natural background sources. The challenge of detecting fast neutrons is twofold. First, the neutron flux from SNM are only a fraction of the corresponding gamma-ray flux. Second, fast neutrons can be difficult to differentiate from gamma rays. The ability to discriminate gamma rays from neutrons combined with neutron imaging can yield large benefit to isolate the localized SNM neutron source from background. With the recent developments of pulse-shape-sensitive plastic scintillators that offer excellent gamma-ray/neutron discrimination, and arrays of silicon photomultipliers combined with highly scalable and fast positron-emission-tomography (PET) multi-channel readout systems, field-deployable neutron imagers suitable for SNM detection might now be within reach. 
In this paper, we present a characterization of the performance of a %
recently available commercial PET-scanner readout, including its sensitivity  to pulse-shape differences between fast neutrons and gamma rays, energy and timing resolution, as well as linearity and dynamic range. 
We find that, while the pulse-shape discrimination is achievable with stilbene, further improvement of the readout is required to achieve it with the best available plastic scintillators. The time and energy resolution appear to be adequate for neutron imaging in some circumstances. %

\end{abstract}


\end{frontmatter}

\section{Single-volume neutron-scatter imagers}

Worldwide, there are now multiple efforts to develop a ``single-volume'' directional neutron detectors~\cite{ Goldsmith2016, Brubaker2018, Jocher2019, Pang2019, Galindo-Tellez:2021ddi, GIHA2021165676, WONDERS2021165294,Guo2021, Zhang:2021kmg}, motivated primarily by a desire to reduce size and complexity of neutron imagers.
An effective way to achieve the detection and characterization of multiple neutron scatters is to use an array of independent detector segments. Better resolution can be achieved by using smaller segments, since the direction uncertainty depends on the interaction position uncertainty, and hence it is inversely proportional to the segment dimensions. The following summarizes some of the key desirable characteristics of a neutron scatter camera: 
\begin{itemize}
    \item Small and mobile%
    \item Safe to operate (no liquids or gases under pressure)
    \item Discriminates between gamma rays and neutrons
    \item Uses a scalable, multi-channel, low-power electronic readout
    \item Uses a scalable and inexpensive detection medium
\end{itemize}

The recent availability of new multi-channel optical sensors in the form of silicon photomultiplier (SiPM) arrays~\cite{ACERBI201916} and 
pulse-shape discrimination (PSD)-capable plastic scintillators~\cite{ZAITSEVA201288, ZAITSEVA2013747, MABE201680} and organic glass scintillators~\cite{CARLSON2016152} are important developments towards the realization of these detectors.
One key missing component however, is a high-channel-density, lightweight, low-power electronic readout capable of simultaneously measuring the particle energy and PSD. 
For example, in our previous setup, like in many multi-channel systems, we used VME-based full-waveform digitizers mounted in a crate~\cite{Li:2019sof, Sutanto:2021xpo}, a solution which was heavy and with high-power consumption. 
The important characteristics of an electronic data acquisition system include:

\begin{itemize}
    \item Good energy resolution over a large dynamic range
    \item Fast sub-nanosecond timing (depending on the detector active volume) %
    to obtain neutron energy from time of flight, needed for neutron directionality
    \item Fine position resolution on neutron interaction vertices
    \item Preserving pulse-shape information %
\end{itemize}

\begin{figure*}
\includegraphics[width=1.0\textwidth]{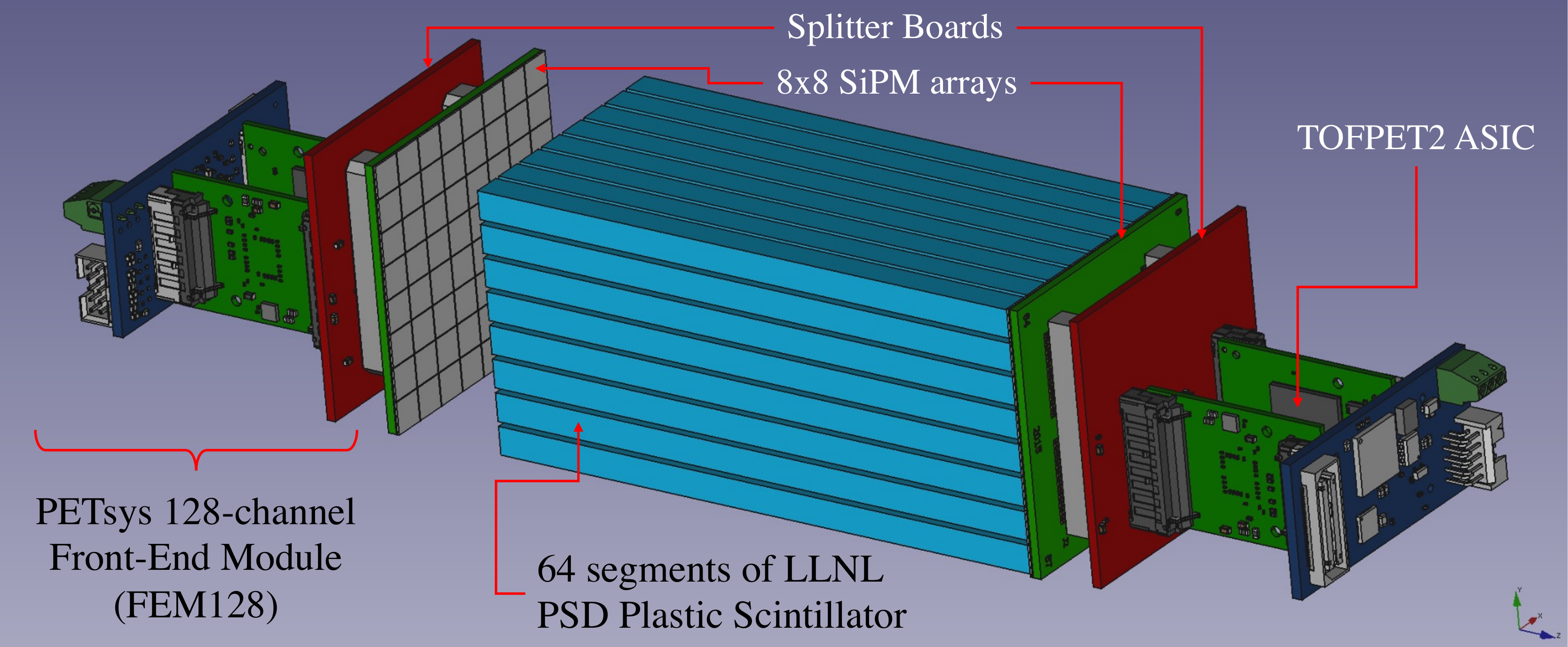}
\caption{Rendering of the detector module. The 8$\times$8 array of $^6$Li-doped PSD plastic rods (5.4~mm $\times$ 5.4~mm $\times$ 11~cm each) with $\sim$1~mm spacing in between. Two 8$\times$8-pixel SiPM arrays are coupled to the scintillator via optical grease. At the back of each SiPM arrays is a splitter board (shown in red), two carrier boards, each with a 64-channel TOFPET2c ASIC. Guiding support frames and ribbon cables are not shown for clarity.}
\label{fig_CAD}
\end{figure*}

The rapid development of specialized technologies for positron emission tomography (PET), which require fast timing over vast arrays of channels, has reduced the cost to
$\mathcal{O}$(\$10)/channel with a time resolution of a few-hundred picoseconds~\cite{Bugalho:2018ptn, Schug:2018klm}.
In the context of neutron imaging, one important question however, 
is whether readouts of this type can preserve a sufficient amount of pulse shape information.

\section{Technical Approach and Performance}

For this  study, we chose an application-specific integrated circuit (ASIC)-based readout system newly developed by the PETsys Electronics corporation for PET applications. %
Unlike the full-waveform readouts, the PETsys readout only outputs the charge and time of each interaction, colloquially referred to as \textit{Qs} and \textit{Ts}. 
Therefore, 
a fully digitized waveform is unavailable for digital pulse processing. This limitation can be addressed by 
splitting the signal from each SiPM pixel into two independent readout channels with different integration times. 
In this way, some of the information about the pulse shape can be preserved. 
The two split channels are referred to as Head and Total,  %
where Head is integrated over a shorter period ($\sim$100~ns), while Total is integrated over a longer period ($\sim$400~ns). 
Particle interaction types are identified on the basis of the ratio of the Head and the Total signal for each pixel. Unlike waveform digitizers, where the integration times can be optimized after digitizing the waveforms, the integration bounds of these PET-based systems must be set at the trigger level (before data collection).

\begin{figure}[ht]%
  \centering
  \includegraphics[width=1.0\linewidth]{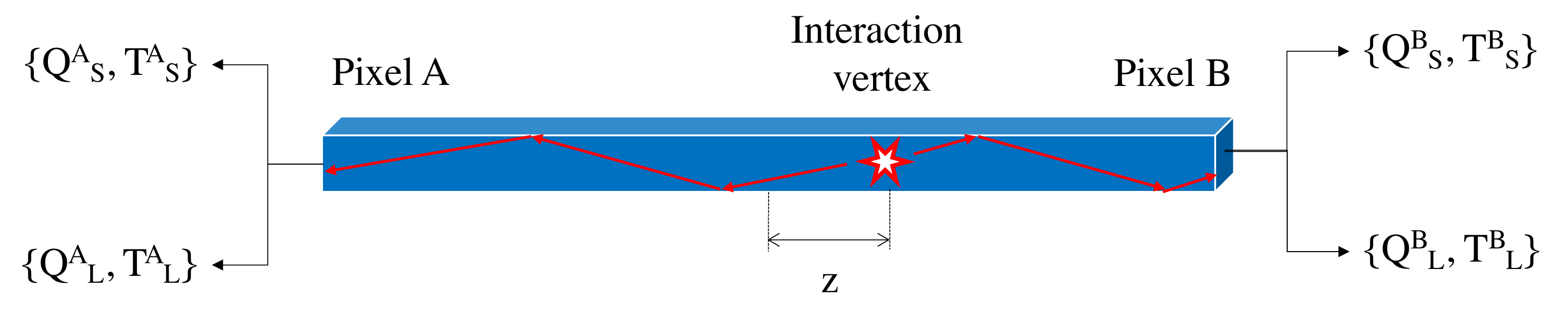}
  \caption{The principle of signal splitting for a single scintillator rod. The scintillation light propagates along the rod (along the $z$-axis) via total internal reflection. SiPM pixels on both ends of the rod transform the light pulse into an electrical signal. A signal from a single SiPM pixel is split into two independent ASIC channels with different integration times: short (``Head'') and long (``Total''). Each readout channel reports charge and timestamp. 
  The $z$ coordinate of where the scintillation light originates is reconstructed with respect to the center of the rod, either by using the ratio of Total charges or timestamp differences on opposite sides of the rod.} %
  \label{fig_splitting_diagram_TIR_2}
 \end{figure}

\begin{figure}%
  \centering
  \includegraphics[width=1.0\linewidth]{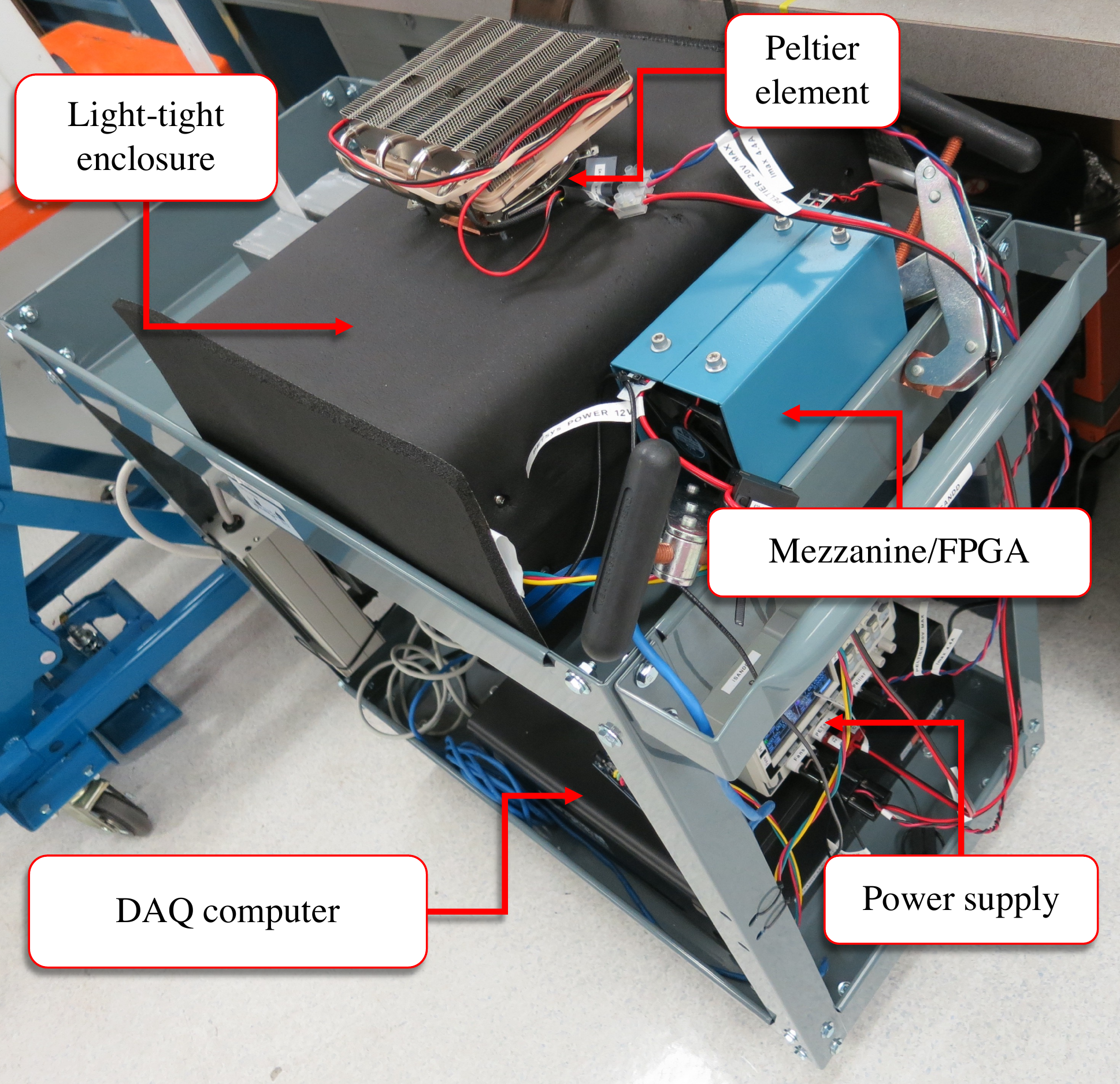}
  \caption{The entire system is placed on a cart. The DAQ computer and power supply are located at the bottom shelf of the cart. The light-tight temperature-controlled enclosure and the mezzanine card (blue box) are situated at the top of the cart.}
  \label{fig_cart}
 \end{figure}

To test the PETsys readout, we constructed a prototype %
system 
referred to in the following as the intelligent Segmented Autonomous Neutron Directional Detector (iSANDD). The iSANDD  consists of sixty-four $^6$Li-doped PSD plastic scintillator rods, each of dimension 5.4\;mm $\times$ 5.4\;mm $\times$ 110\;mm, coupled to two 8$\times$8-pixel Sensl J-60035 SiPM arrays. The readout of each SiPM is mounted at the back. A rendering of the primary detection module is shown in Fig.~\ref{fig_CAD}.
In the following, we define the axes $x$ and $y$ as being in the plane of SiPM array face. The $z$-axis runs parallel to the scintillator rods.
The signals from each pixel are split into two (red board) and then digitized into $T$ and $Q$ values  using the green TOFPET2c boards. The $T$ and $Q$ values are then transmitted digitally via a ribbon cable to a mezzanine board (not shown). 
In our data analysis, we require the charge $Q$ and timestamp $T$ to be available on both ends of a single rod, as illustrated in Fig.~\ref{fig_splitting_diagram_TIR_2}. The $z$-coordinate is defined as the distance along the $z$-axis from the center of each rod. 
The mezzanine board is shown in Fig.~\ref{fig_cart} inside the blue box. The mezzanine board accepts up to 1024 channels from up to 8 ribbon cables. It incorporates the main clock so that all 1024 channels can be assigned global timestamps. It also has an FPGA, which enables the user to set trigger, gain and integration parameters. The SiPM voltage is set via the PETsys software.

\begin{figure}%
  \centering
  \includegraphics[width=1.0\linewidth]{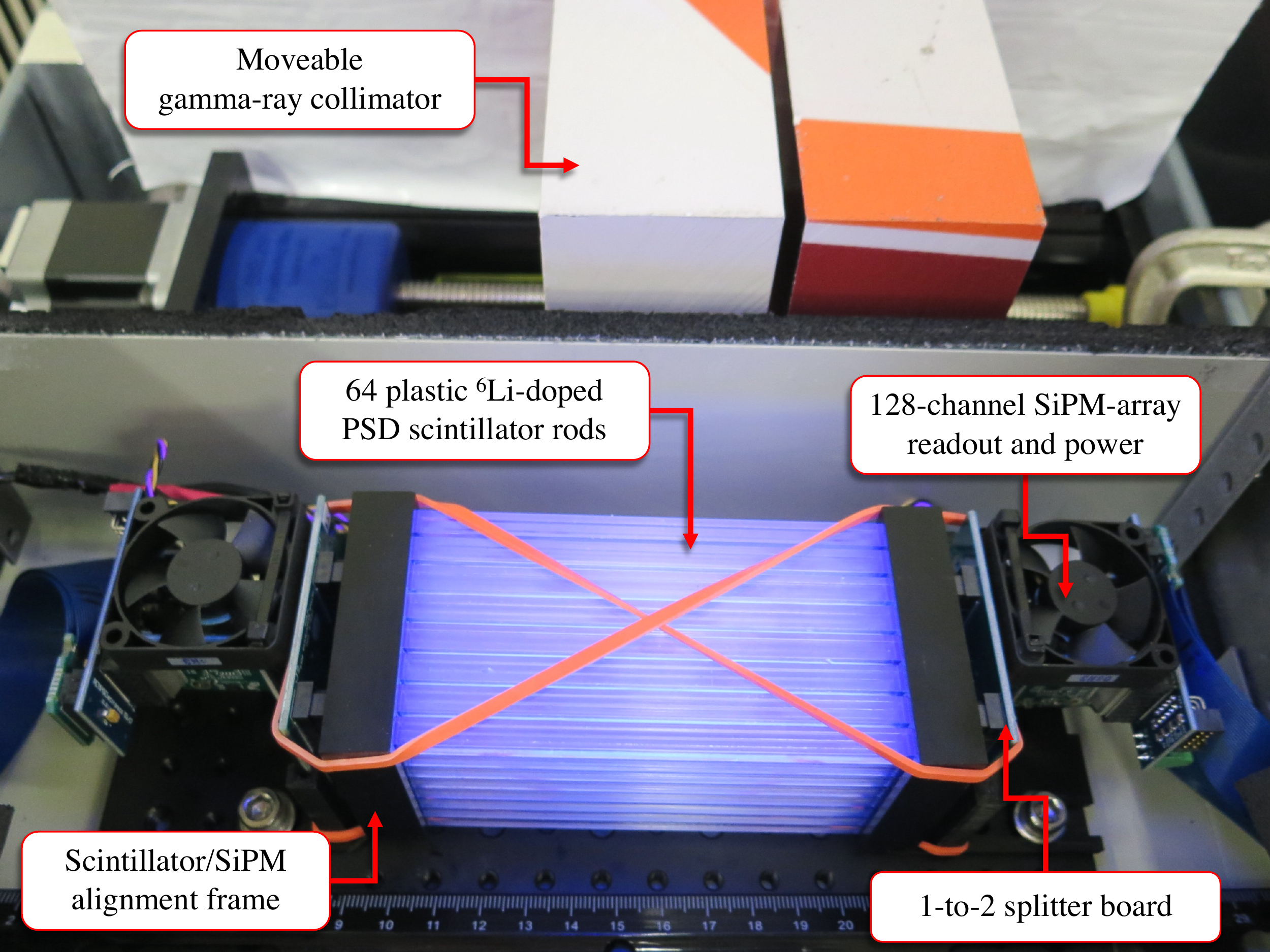}
  \caption{iSANDD module incorporating 64 $^6$Li-doped (0.1\% wt) plastic PSD scintillator rods ($110 \times 5.4 \times 5.4$~mm$^3$) with SiPM arrays and compact electronics readout. A lead collimator is placed on an motorized translation stage to calibrate the detector response}
  \label{fig_lead_setup}
 \end{figure}

One of the complications associated with the use of this system is its sensitivity to temperature variations. We constructed a dark box covered with thermally-insulating foam to stabilize the temperature. A Peltier element and a radiator with a fan are mounted on the top of dark box. The Peltier element voltage is set at 20\;V and the current draw is 2\;A.  We also equipped each ASIC with a heat sink and a fan mounted on top of the FEM128, as shwon in Fig.~\ref{fig_lead_setup}. 
In our tests, the ASIC temperatures normally settled at about 30\;$^\circ$C. %
An interlock system was developed to automatically shut down the system if the temperature on any ASIC rises above 35\;$^\circ$C. %
Additional humidity and temperature sensors were also mounted within the dark box. 
Having a humidity sensor is particularly important to ensure that there is no unwanted condensation inside the light-tight enclosure due to the Peltier element.
Two connections were provided to the mezzanine board: 12-V power and Ethernet, which provides connection to a Linux desktop. The nominal power draw of the mezzanine alone is 12~W; each FEM128 adds 3~W. For our 256-channel system, the power draw was 18~W. 
For a short mobility test we successfully powered and operated the electronics readout, including the SiPM bias and cooling fans but not including the Peltier element, by a 12-V battery pack.
The desktop computer has 3 Ethernet interfaces: internal network, communication with PETsys, and communication with the power supply. The entire 
system, including the power supply and the DAQ computer, is mobile and compact, and was placed on a small hand-cart, as shown in Fig.~\ref{fig_cart}.

To characterize each interaction, the charge and relative timing of signals were recorded. Each pixel signal is split into two readout channels, which are used to record the integration times for Head and Total charge.
To determine the event position within a scintillator rod, the ratio of Total charge recorded at both ends of the rod
is used. Since both the Head and Total are needed to determine the PSD, the minimum requirement for physics events, therefore, is a Total and Head trigger on both ends of a single rod. In the offline analysis, the requirement is therefore four channels (per rod) for an event to be considered a physics event.
There is some tuning of the data acquisition parameter settings required before physics data can be taken. For example, the Head and Total integration times need to be set before hand. 
The SiPM operating voltage was set to 30~V, and the electronics gain was set to the highest setting.
PETsys uses three different acquisition modes: QDC, TOT, and mixed~\cite{Bugalho:2018ptn}.
The TOT mode records the time over the threshold.
The QDC mode records an integrated charge for a preset time window.
In most tests described in this study, we used the QDC mode and the same thresholds for all Head and Total channels. %
The integration time windows were set to approximately 100~ns for Head and 400~ns for Total.

\begin{figure}[ht]
\begin{multicols}{2}
\begin{centering}
    SiPM A \par 
    SiPM B \par 
\end{centering}
\end{multicols}
  \vspace{-3mm}
\centering
  \includegraphics[width=0.49\linewidth]{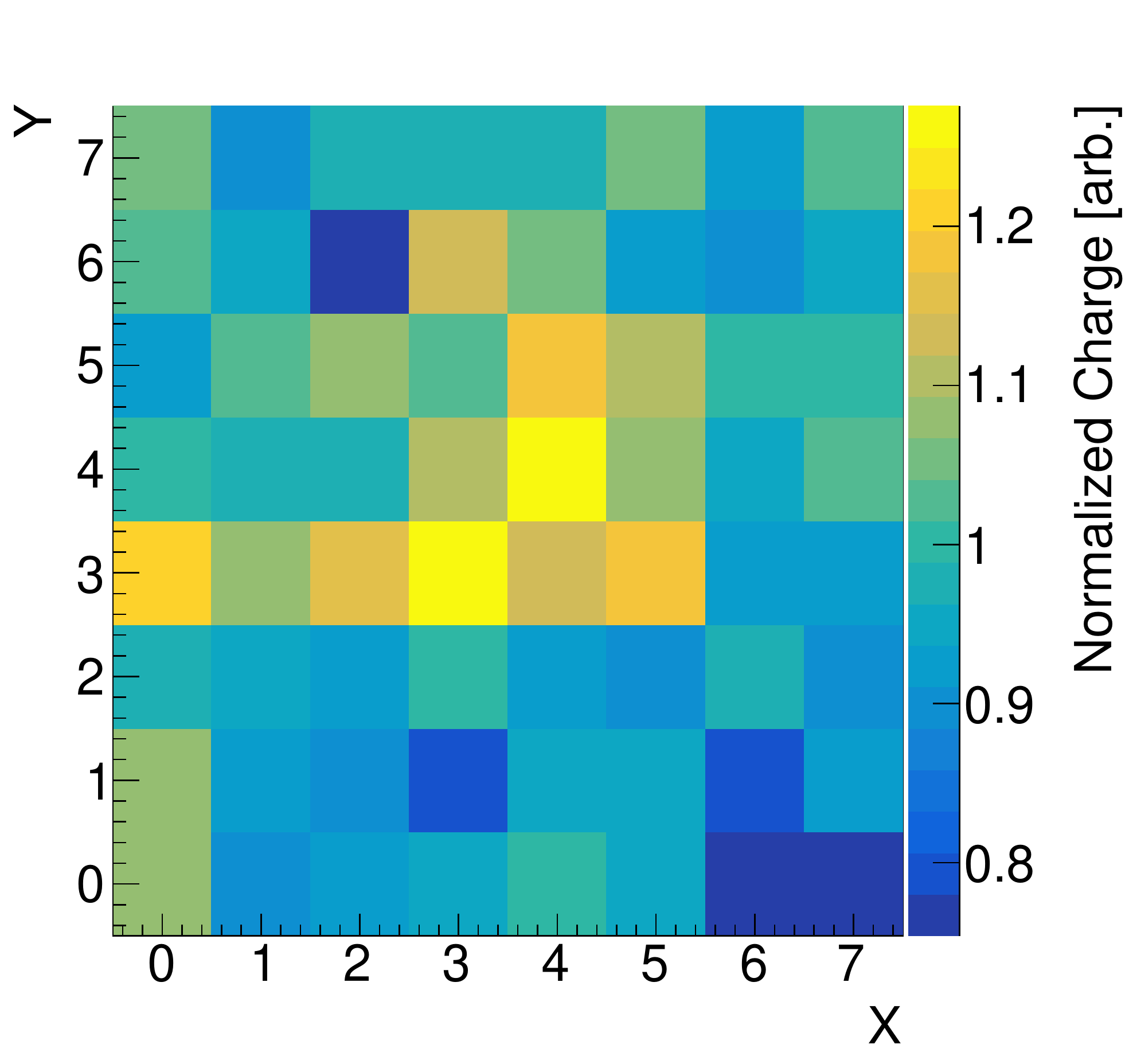} %
  \includegraphics[width=0.49\linewidth]{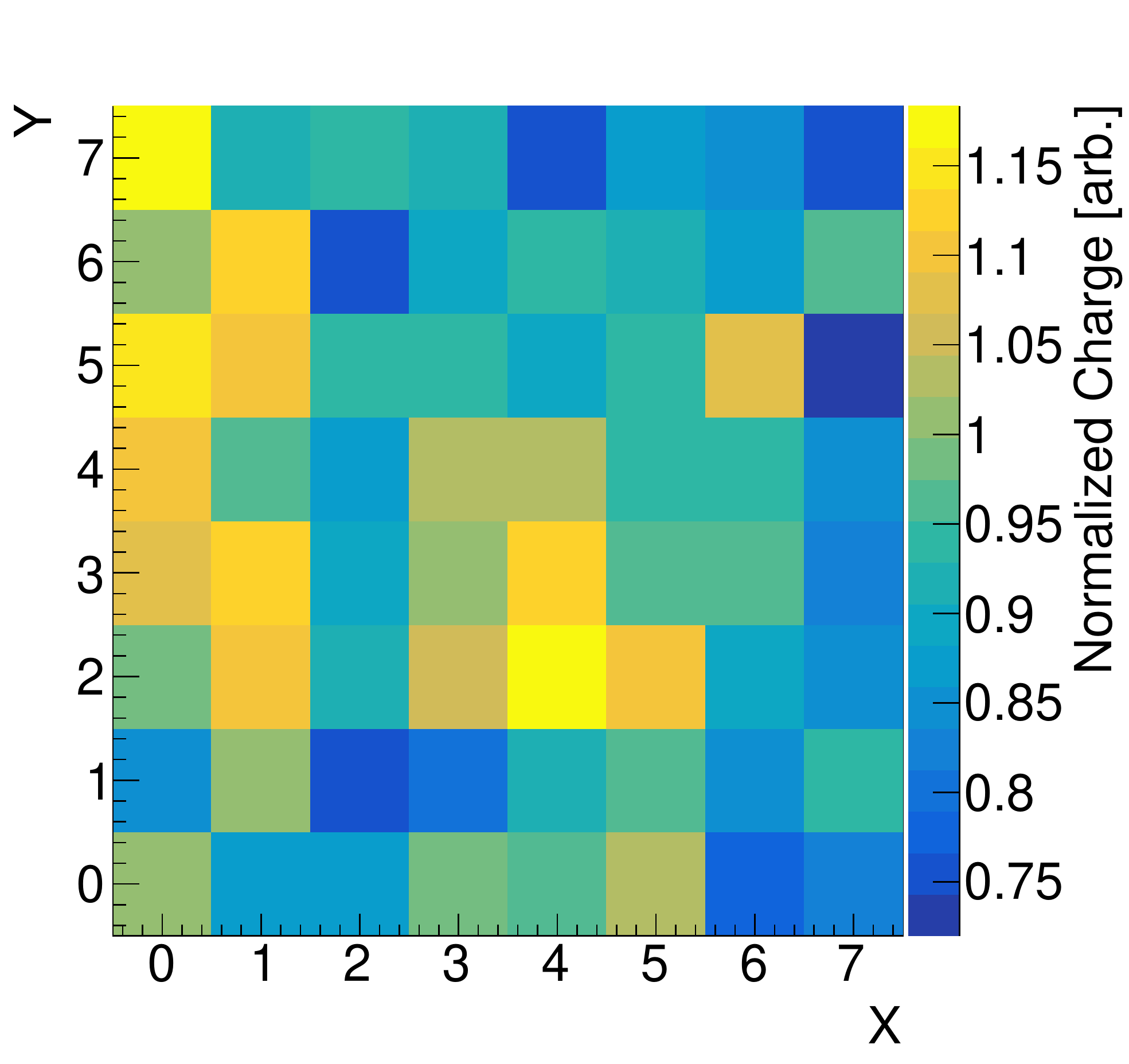} %
    \caption{SiPM response with scintillator rods coupled. The collimated $^{137}$Cs gamma-ray fan-like beam is oriented at the center of the rods. The pixel responses are normalized to pixel (3, 3) on SiPM B.}
    \label{fig_Cs_flat_fielding}
\end{figure}

\begin{figure*}[ht!]
\begin{multicols}{3}
\begin{centering}
    $^{137}$Cs \par 
    $^{54}$Mn \par 
    $^{22}$Na \par 
\end{centering}
\end{multicols}
\vspace{-.9cm}
\begin{multicols}{3}
    \includegraphics[width=1.0\linewidth]{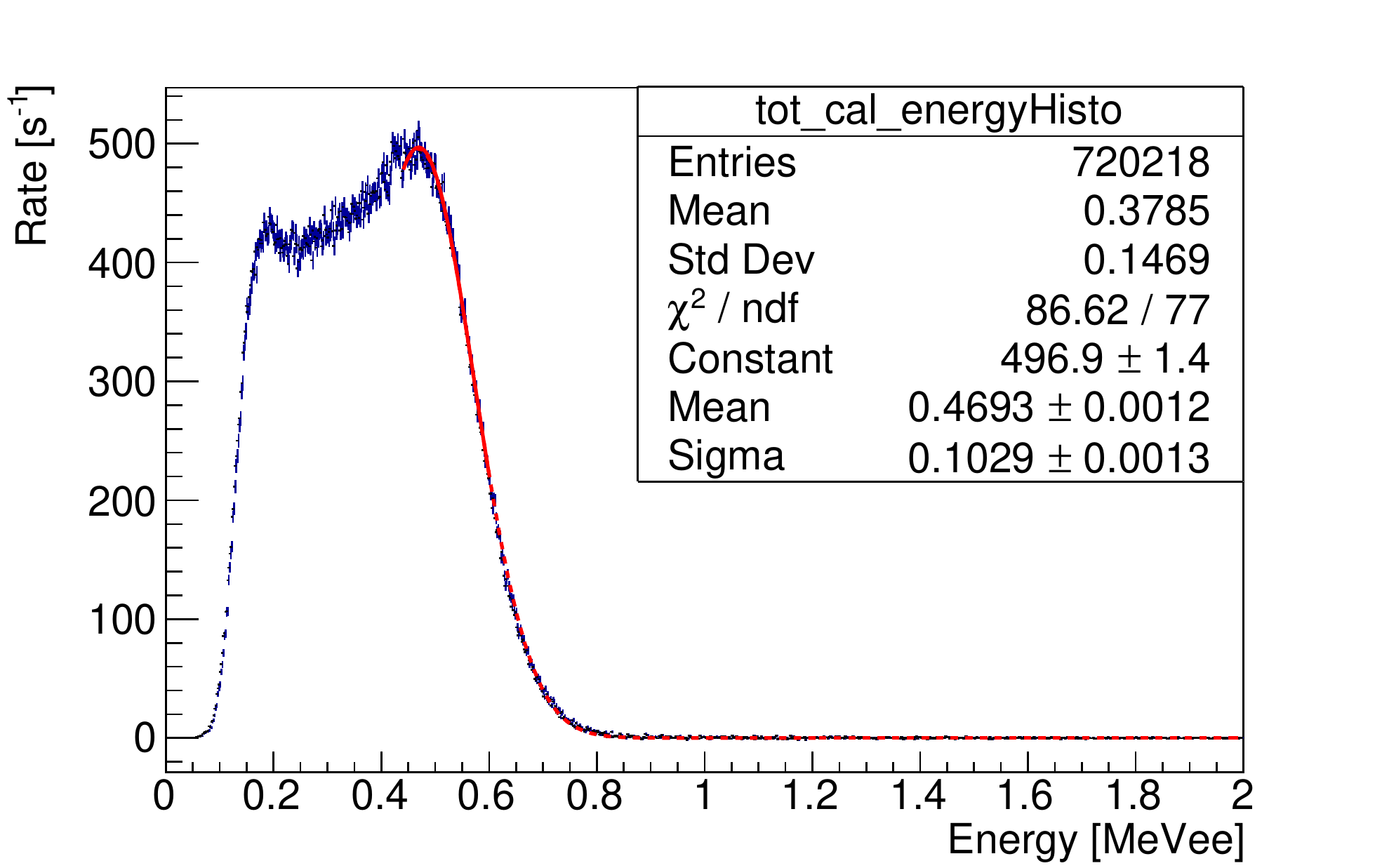} \par 
    \includegraphics[width=1.0\linewidth]{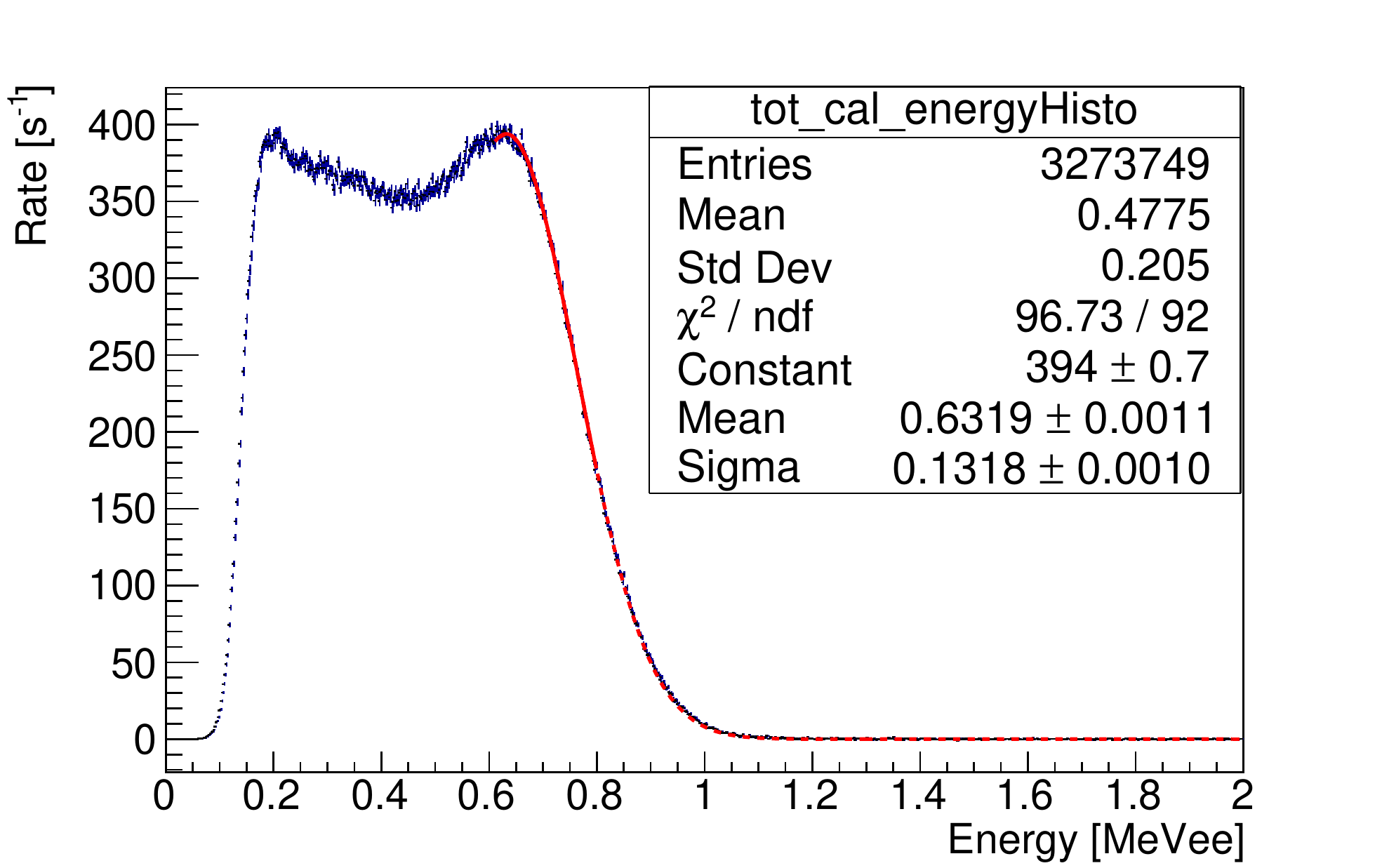} \par 
    \includegraphics[width=1.0\linewidth]{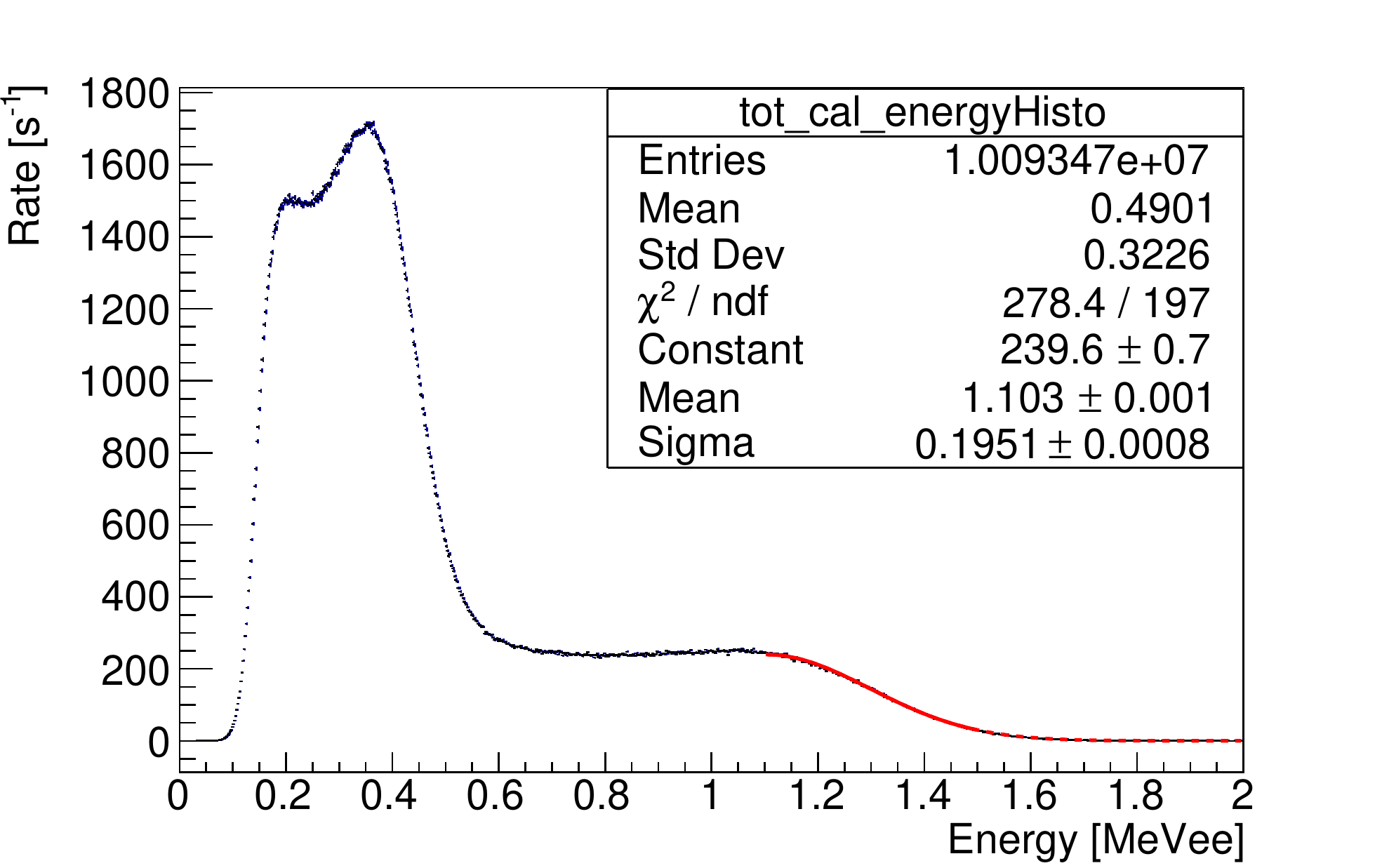} \par 
\end{multicols}
\vspace{-0.5cm}
\caption{Gamma spectra for center-collimated sources: $^{137}$Cs, $^{54}$Mn, and $^{22}$Na. The response is recorded across multiple rods and is background-subtracted.}
\label{fig_gamma_spectra}
\end{figure*}

The scintillator rods are aligned with the SiPM array using a special support frame, shown in Fig.~\ref{fig_lead_setup}. %
Optical grease (EJ-550) is applied 
to ensure minimal refractive index mismatch due to an air gap between the scintillator and the SiPM.
The rods are separated from each other by a $\sim$1-mm air gap to ensure total internal reflection (TIR) within each rod.

\subsection{Energy characterization}

To characterize the energy response and resolution as a function of position along the scintillator rods,
a fan-like beam of gamma rays, was directed perpendicular to the scintillator rods, as shown in Fig.~\ref{fig_lead_setup}. 
The figure shows a motorized translation stage 
used to automate the detector calibration for any $z$-position. The translation stage incorporated a 3D-printed holder to support a lead-brick collimator and a gamma-ray source. 
With the collimator set to the center of the scintillator array, the  response of each of the pixels was measured, as shown in Fig.~\ref{fig_Cs_flat_fielding}. This process is referred to colloquially as ``flat-fielding'', with the ultimate aim of generating a map of normalization factors for all pixels.
The normalization map is used to ensure consistent response across both SiPM arrays. 
Spectra generated from $^{137}$Cs, $^{54}$Mn, and $^{22}$Na are shown in Fig.~\ref{fig_gamma_spectra}, with the collimated beam at the center of the scintillator array.
The mean energy resolution, as measured by a Gaussian fit of the Compton edge  of each of the 64 scintillator rods, was 22\% for $^{137}$Cs. This was somewhat worse than the resolution of the same detector ($\sim$16\%) measured using full-waveform digitizers~\cite{Li:2019sof}.

\begin{figure}[ht]%
  \centering
  \includegraphics[width=0.5\textwidth]{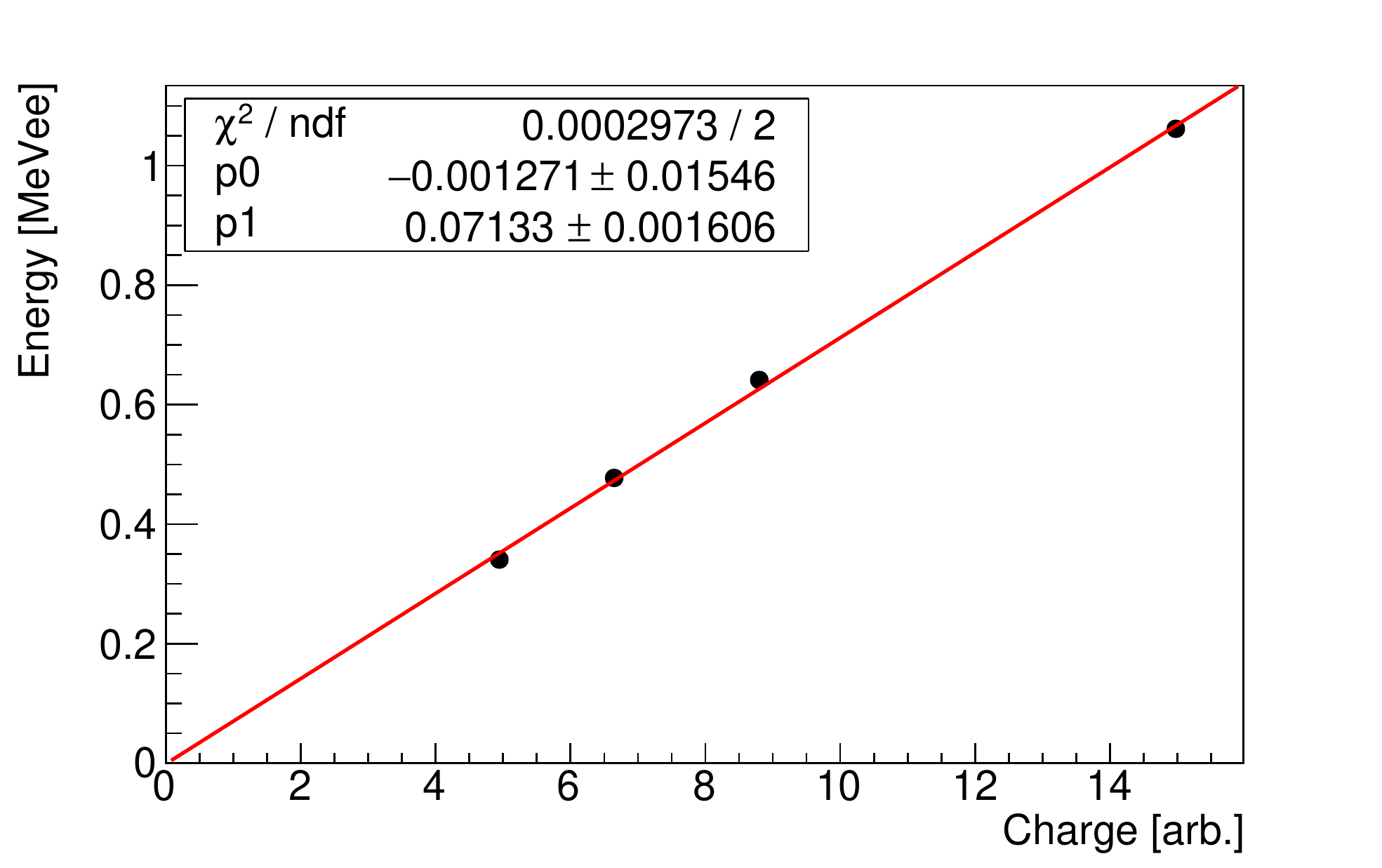}
  \caption{The linearity of the energy response obtained from the summed-rod Compton-edge positions shown in Fig.~\ref{fig_gamma_spectra}}
  \label{fig_gamma_linearity}
 \end{figure}

The linearity of the detector energy response up to 1274~keV is shown in Fig.~\ref{fig_gamma_linearity}.
The line of best fit of the four gamma-ray points intersects the y-axis at $-0.0013 \pm 0.0155$, consistent with the line through the origin.

\begin{figure}[ht]
    \includegraphics[width=1.\linewidth]{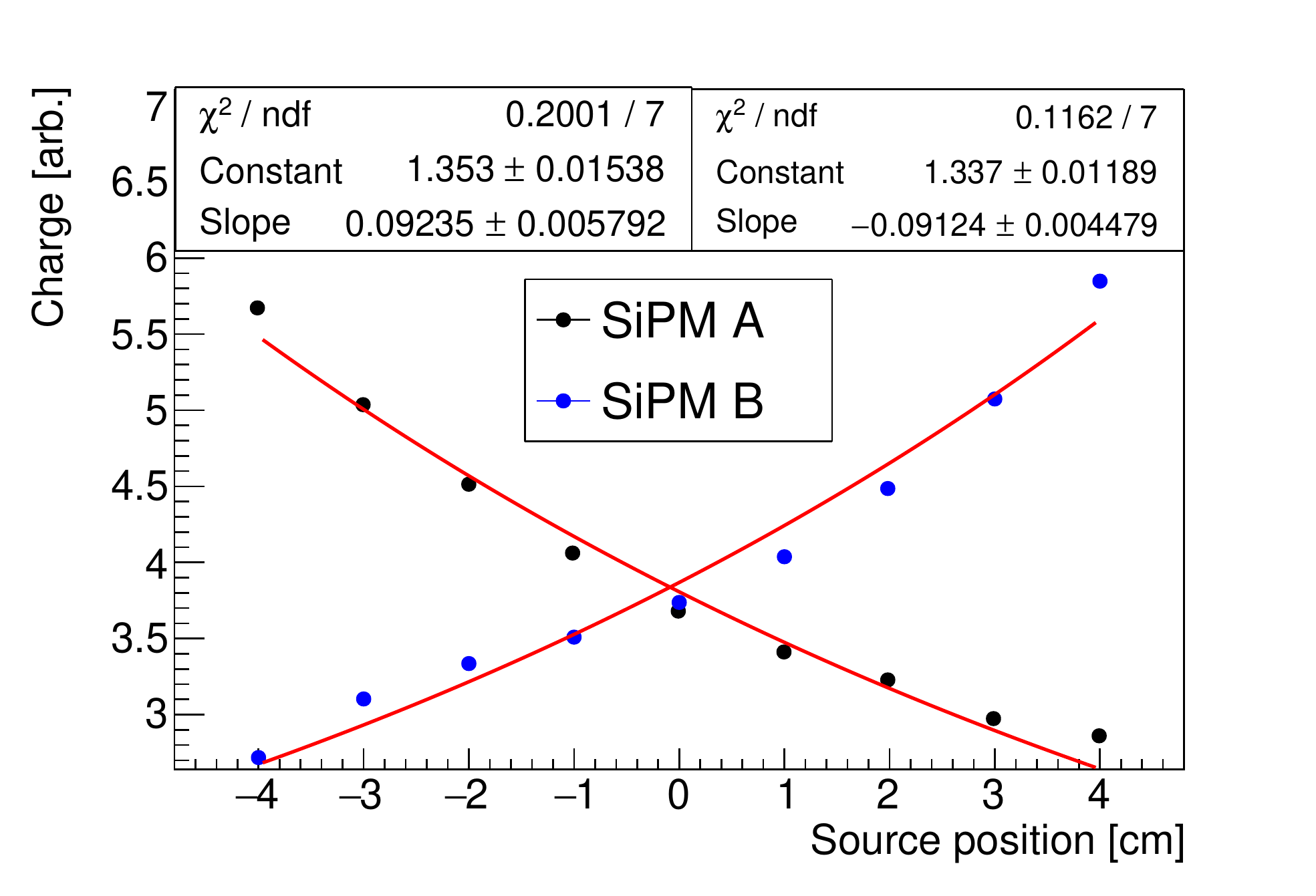}\\ 
\vspace{-0.5cm}
\caption{Attenuation length measurement on SiPM arrays A and B. The collimated gamma-ray source was moved along the rods.}
\label{fig_attL}
\end{figure}

The effective attenuation length along each of the rods was measured 
using a collimated gamma-ray beam at known $z$-positions and the relative signal strength as measured by the pixels at either end of each scintillator rod, as shown in Fig.~\ref{fig_attL}.

\begin{figure}[ht!]
    \includegraphics[width=1.\linewidth]{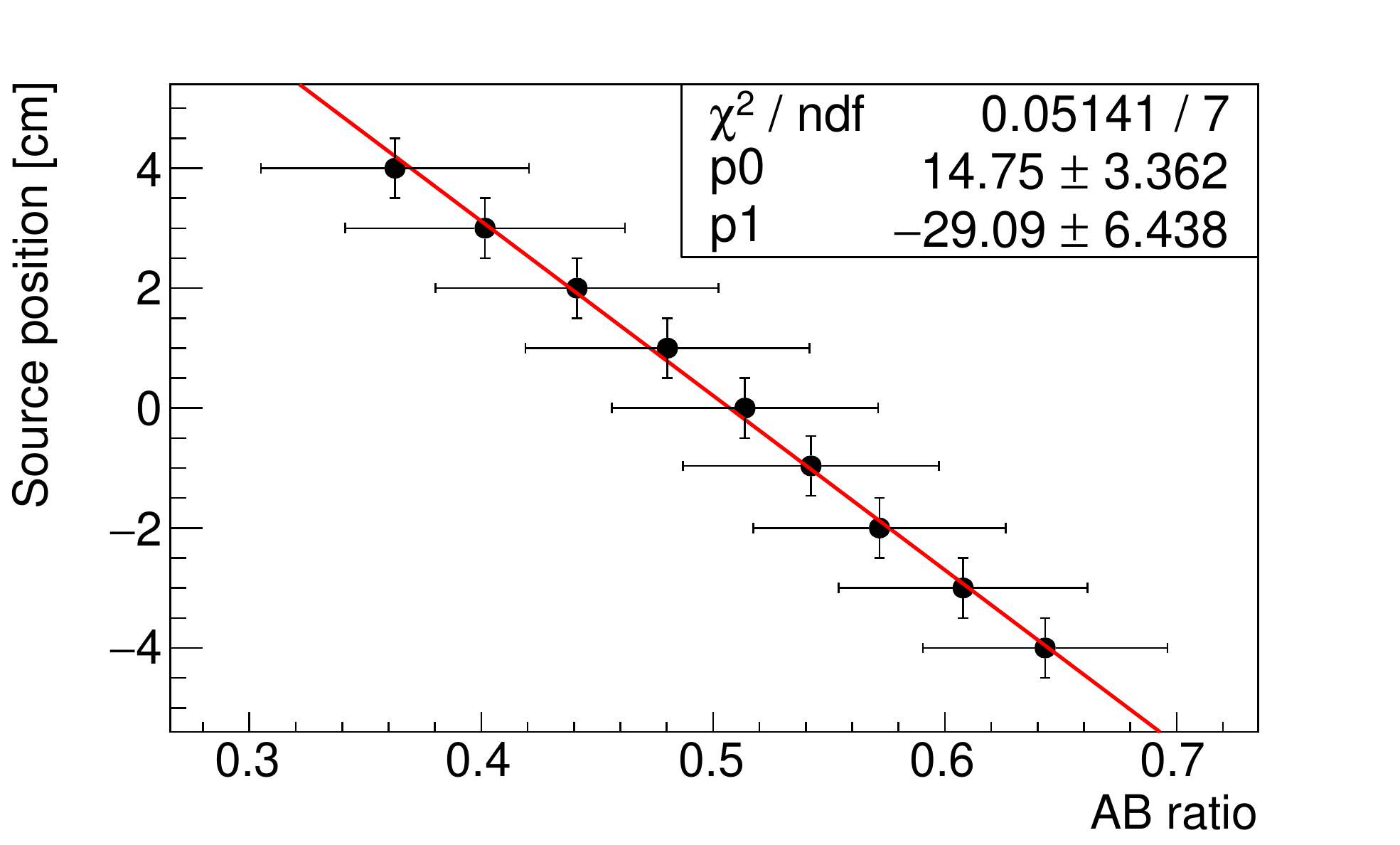} \par 
\vspace{-0.5cm}
\caption{Dependence of position of interaction along the rod on the charge ratio $R_{AB}$. }
\label{fig_diffZ}
\end{figure}

The position uncertainty can be measured using relative charge as measured at both ends of the scintillator rod. 
This ratio is called the ``AB'' ratio and is defined as
\begin{equation}
    R_{AB} = \frac{Q_A}{Q_A + Q_B},
\end{equation}
where $Q_A$ and $Q_B$ are the measured charges at both ends  of each scintillator rod.
The AB ratio as a function of position is shown in Fig.~\ref{fig_diffZ}. The uncertainties shown represent the 1-sigma width of a Gaussian fitted to the AB ratio at each position and, hence, the position uncertainty.

A peculiarity of the PETsys readout was an observation of seemingly random baseline and gain shifts on a run-to-run basis. Our preliminary results indicated that these shifts were aggravated by temperature fluctuations. 
We also operated iSANDD outside of the temperature controlled enclosure to measure the effect of changes in temperature on the energy response. We noticed that the combination of baseline drift and energy response caused as much as 40\% drop in the 511-keV Compton edge ADC value for a 3-degree temperature increase. For 1274~keV, the equivalent temperature change caused a deviation of $\sim$15\%.

\begin{figure}%
  \centering
  \includegraphics[width=1.0\linewidth]{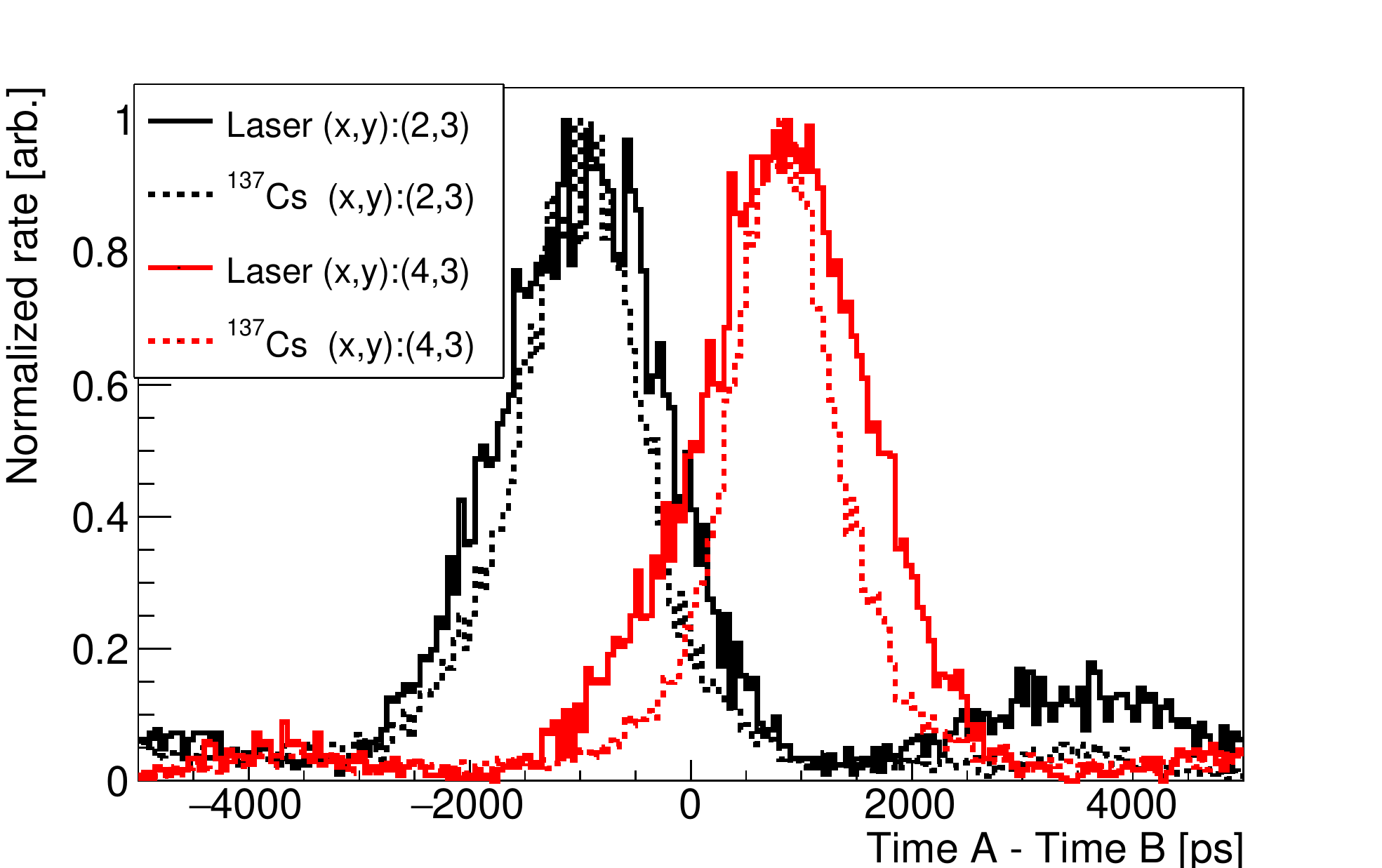} \par 
  \caption{Timing offsets measured between two pairs of pixels using a laser and a center-collimated $^{137}$Cs source show consistency in the mean values and the width of the distributions.}
  \label{fig_timing_offsets_2rods_laser_Cs}
\end{figure}

\begin{figure}%
  \centering
  \includegraphics[width=1.0\linewidth]{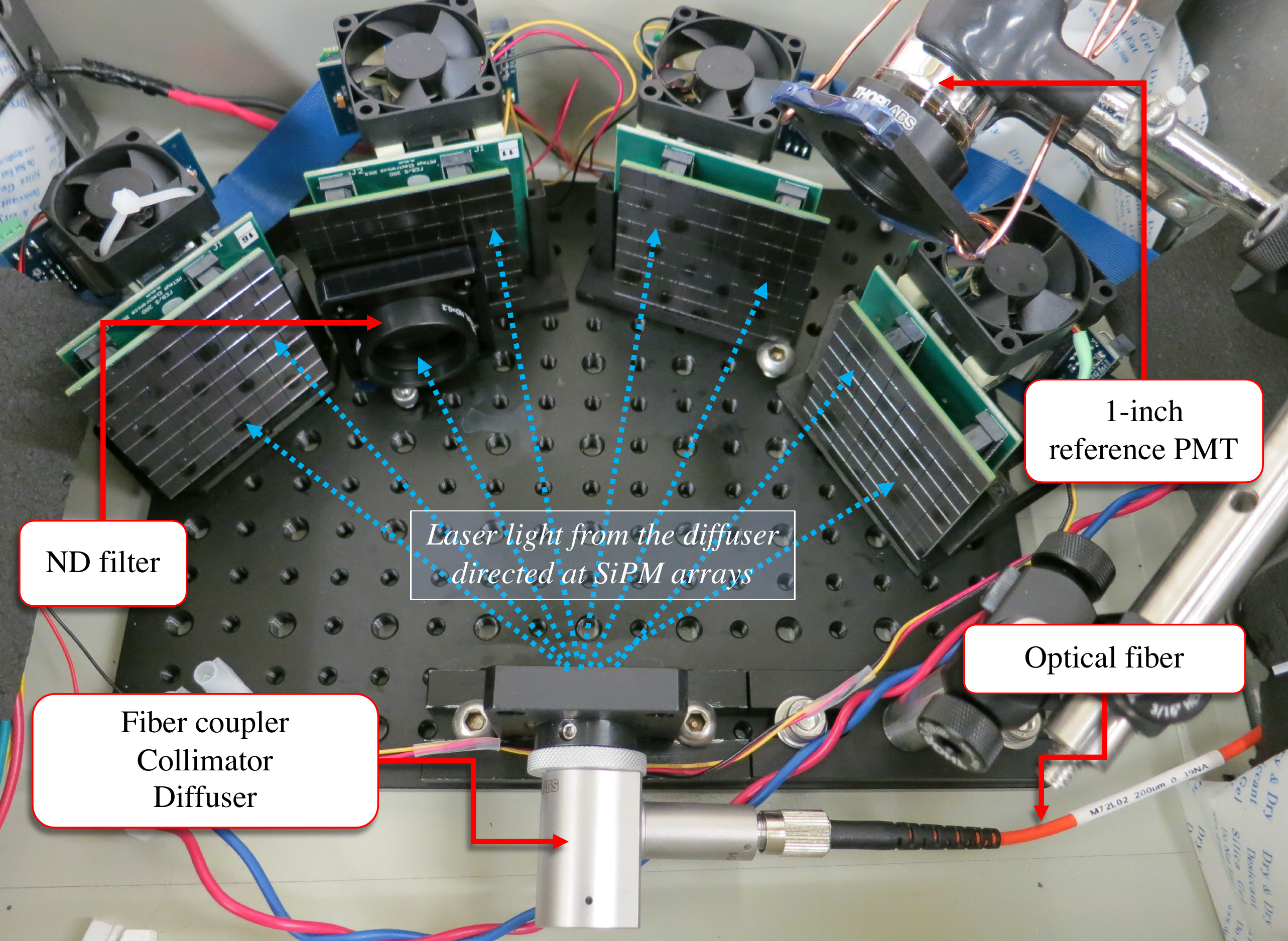}
  \caption{Laser calibration system. The picosecond laser beam is fed into the enclosure via a fiber and collimator. The beam is diffused such that multiple pixels are illuminated to study and calibrate timing response across the SiPM arrays. A reference 1-inch PMT is also shown in the top right corner. A neutral-density filter is shown covering several SiPM pixels to study the timing response to varying relative light intensity between the pixels.}
  \label{fig_laser_setup}
 \end{figure}

\begin{figure}[ht!]%
\begin{multicols}{2}
\begin{centering}
    SiPM A \par 
    SiPM B \par 
\end{centering}
\end{multicols}
  \vspace{-3mm}
  \centering
  \includegraphics[width=.49\linewidth]{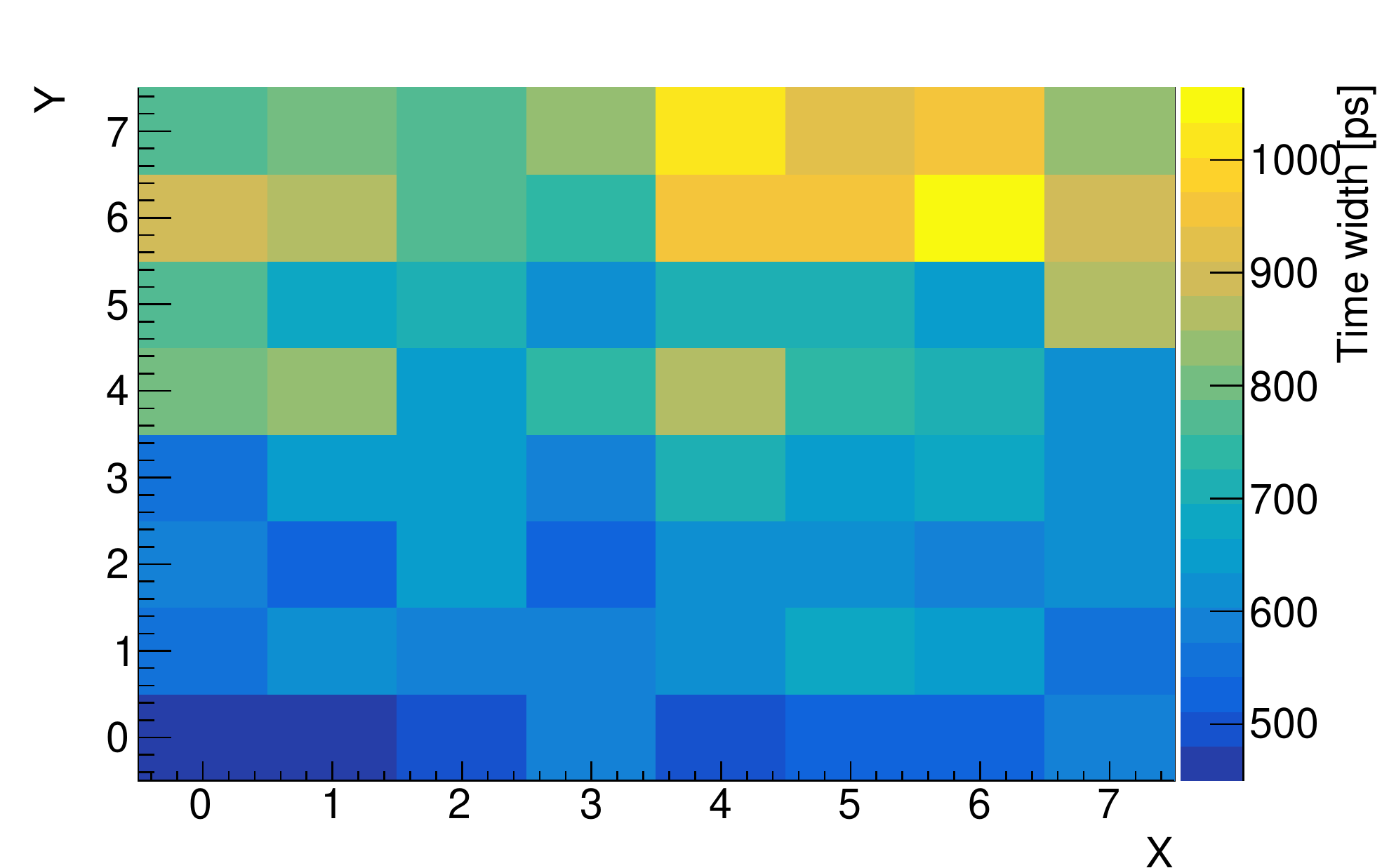}
  \includegraphics[width=.49\linewidth]{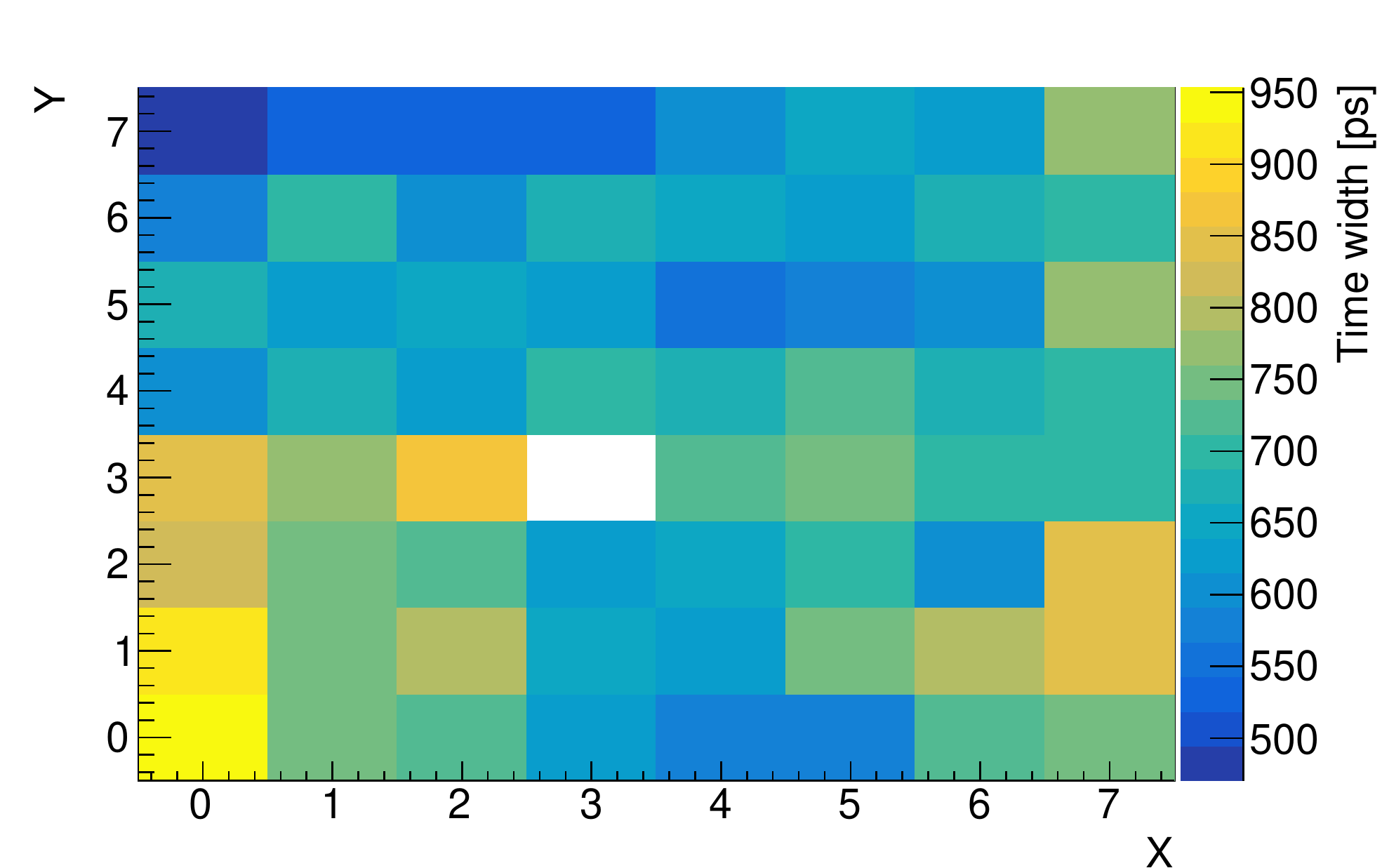}\\
  \includegraphics[width=.49\linewidth]{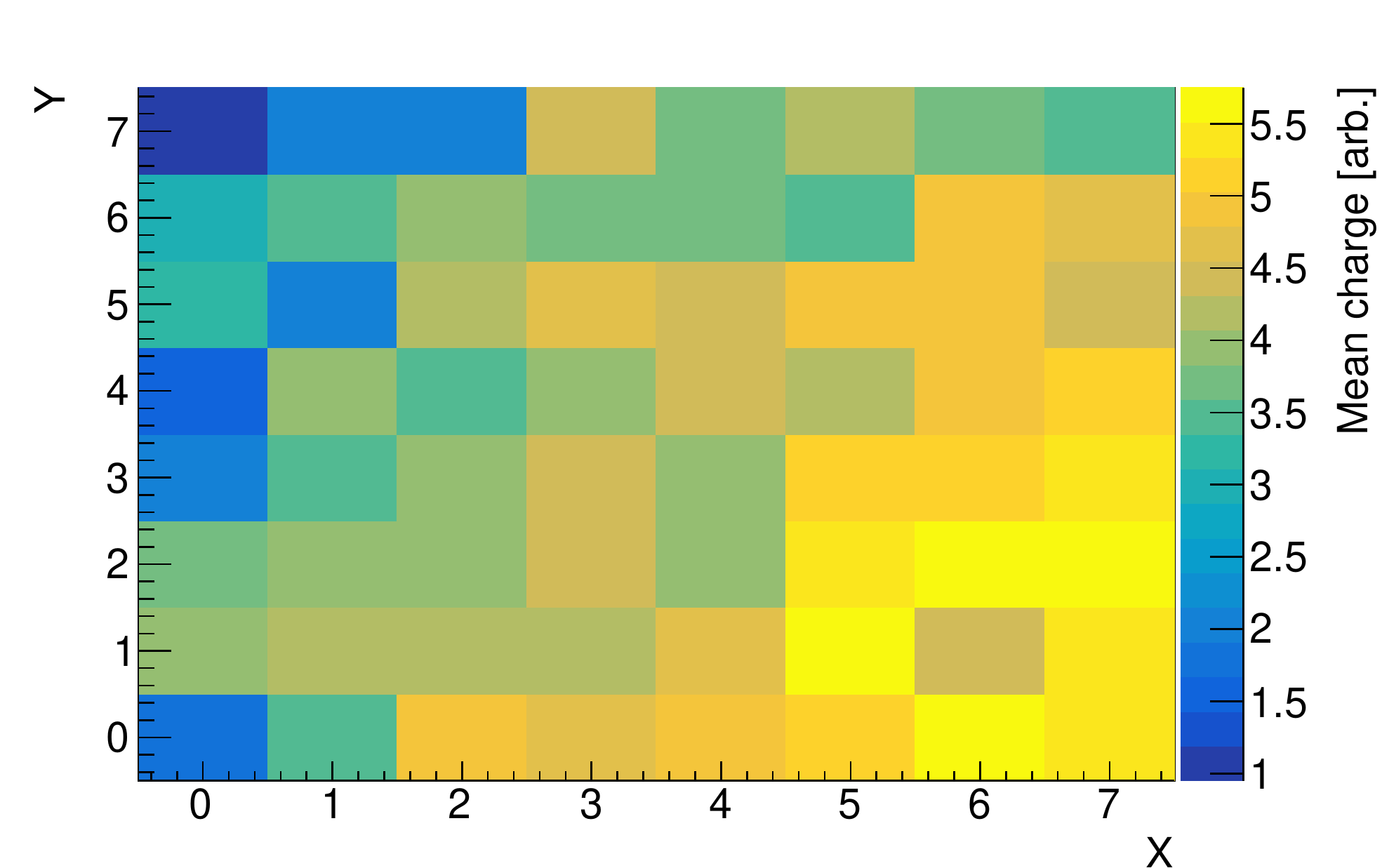}
  \includegraphics[width=.49\linewidth]{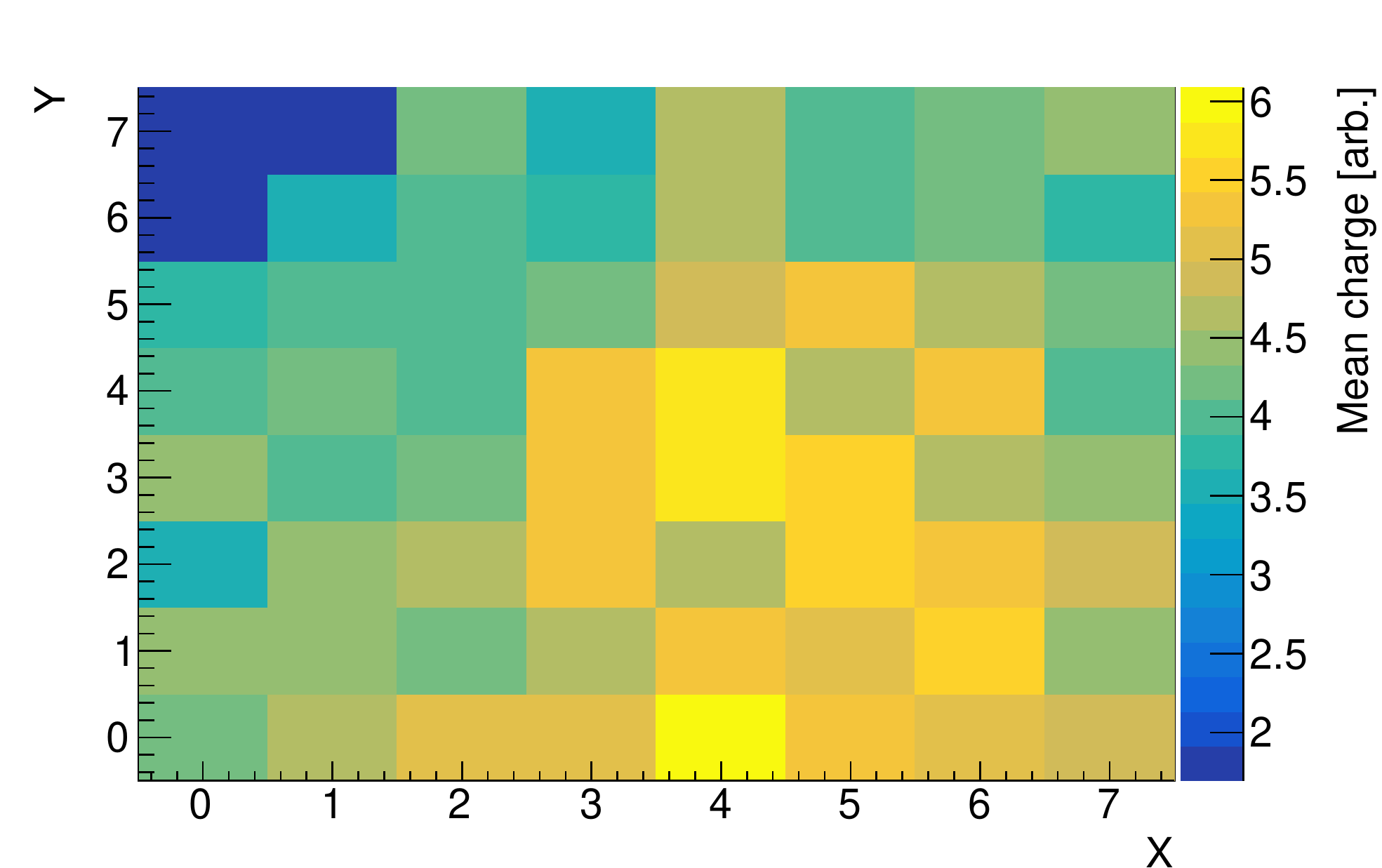}\\
  \includegraphics[width=.49\linewidth]{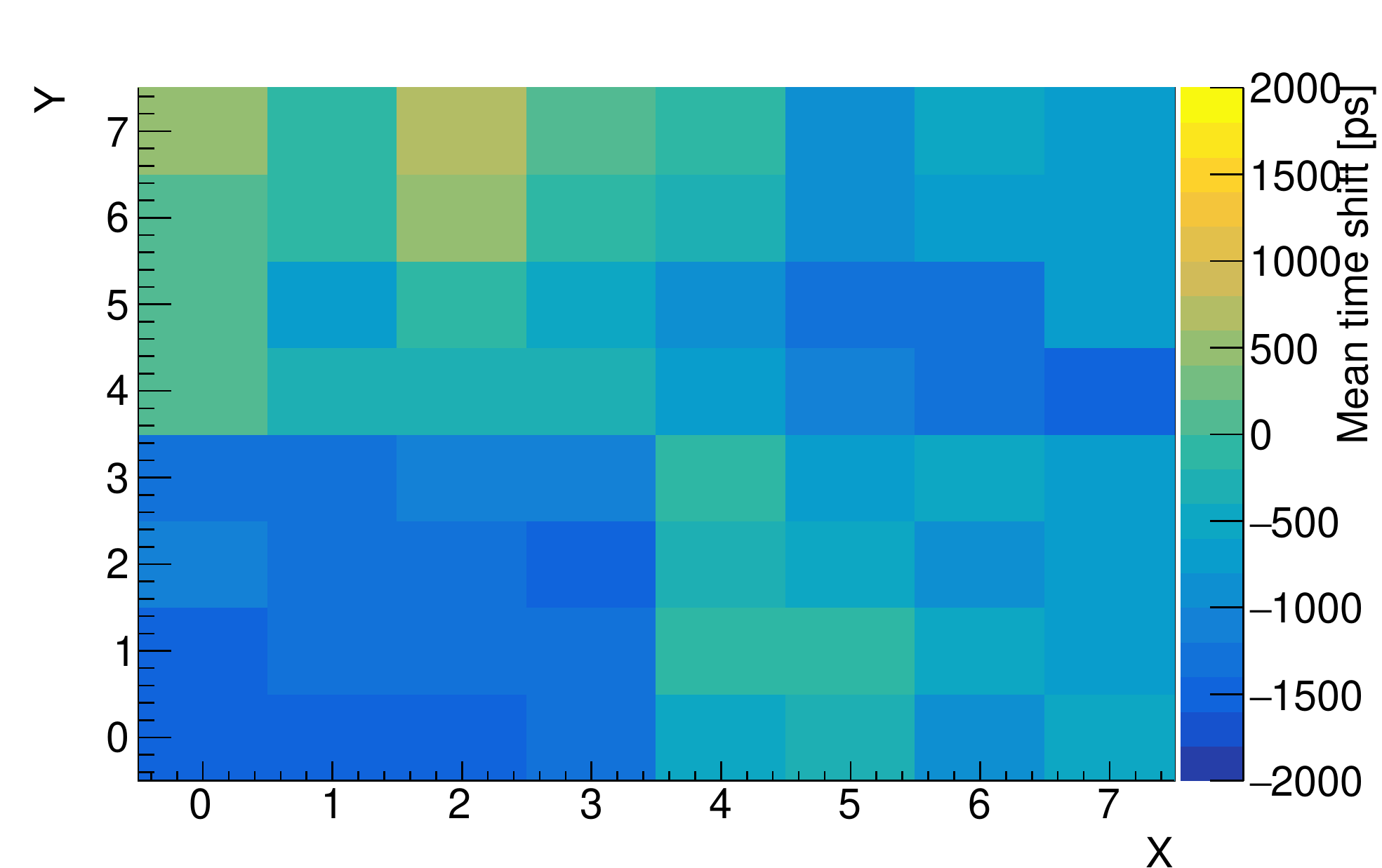}
  \includegraphics[width=.49\linewidth]{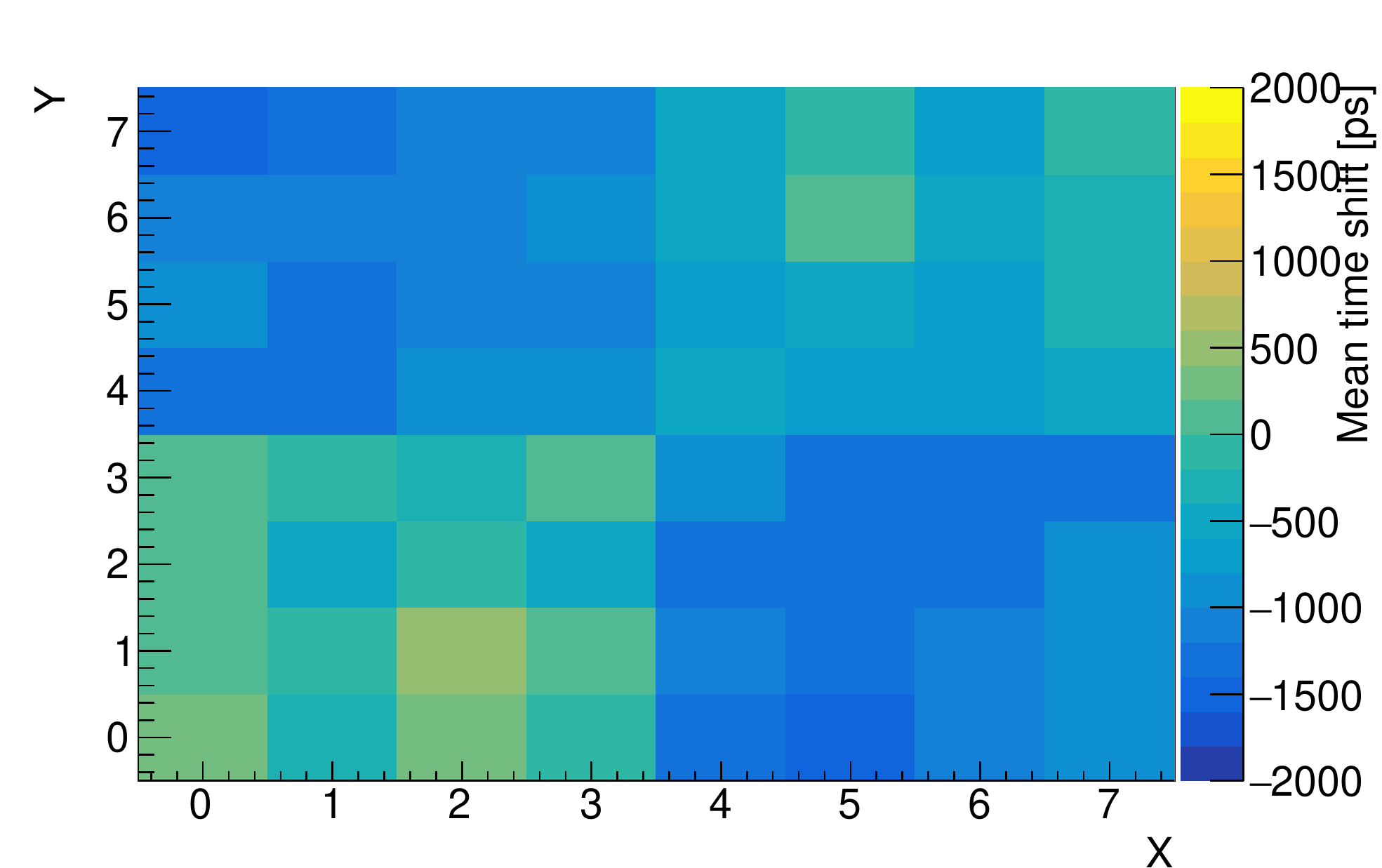}\\

  \caption{Laser calibration measurements. Left column: SiPM A; Right column: SiPM B.
  {\it Top: } Gaussian widths of the timing variations in laser tests for two 8$\times$8 SiPM arrays. 
  {\it Middle: } Mean charge distribution. The nonuniformity of response is primarily due to the nonuniformity of the diffused laser beam, as seen by comparison with Fig.~\ref{fig_Cs_flat_fielding}.
  {\it Bottom: } Mean time shift. %
  A lookup table for correction is constructed based on these measurements. For the ``Time Width'' and ``Mean Time Shift'' plots, the pixel (3,3) on the SiPM array B was used as a reference. 
  }
  \label{fig_T_width_map_A}
 \end{figure}

\begin{figure*}[ht!]
\begin{multicols}{3}
\begin{centering}
    (a) \par 
    (b) \par 
    (c) \par 
\end{centering}
\end{multicols}
\vspace{-.8cm}
\begin{multicols}{3}
    \includegraphics[width=1.0\linewidth]{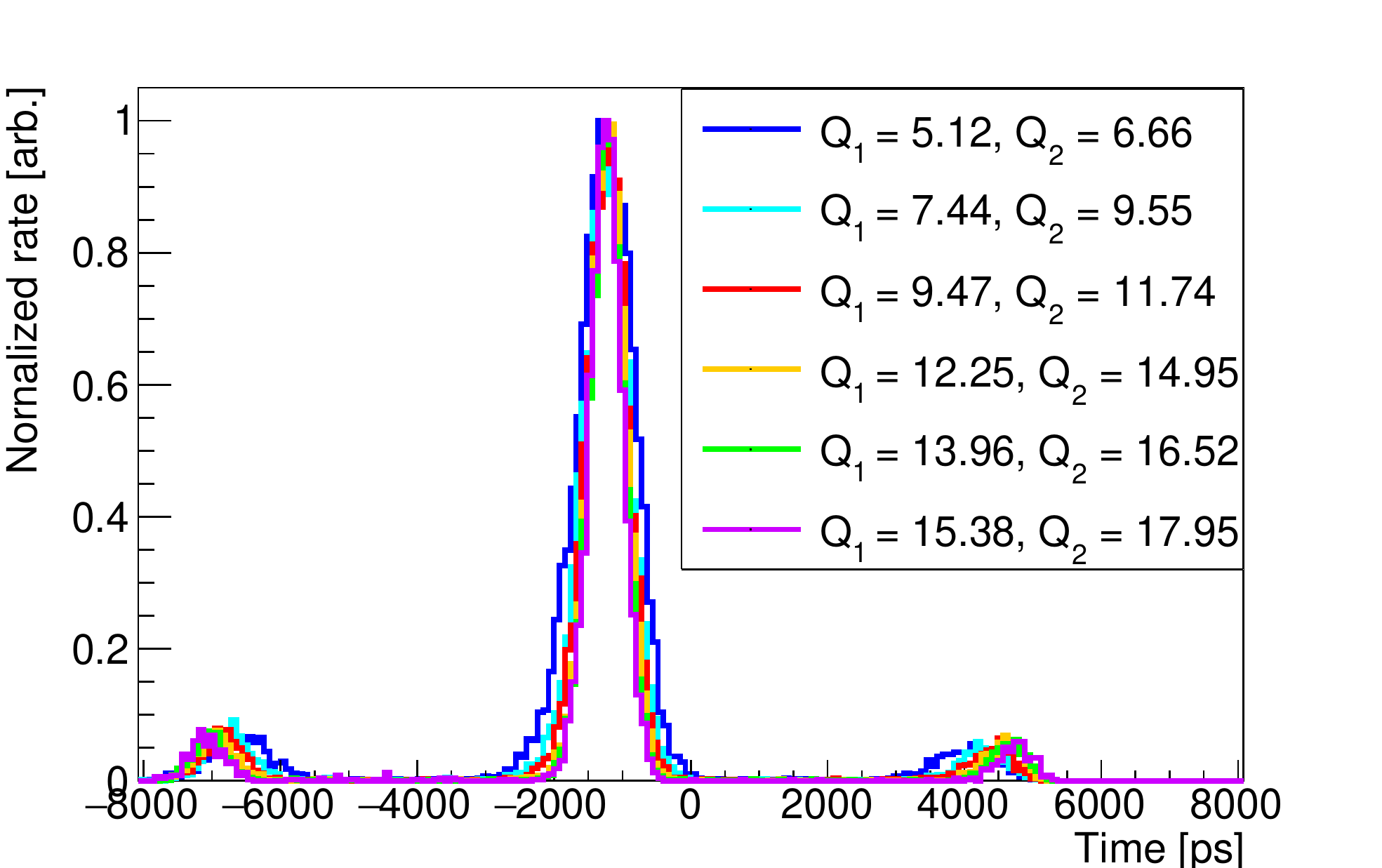} \par %
    \includegraphics[width=.9\linewidth]{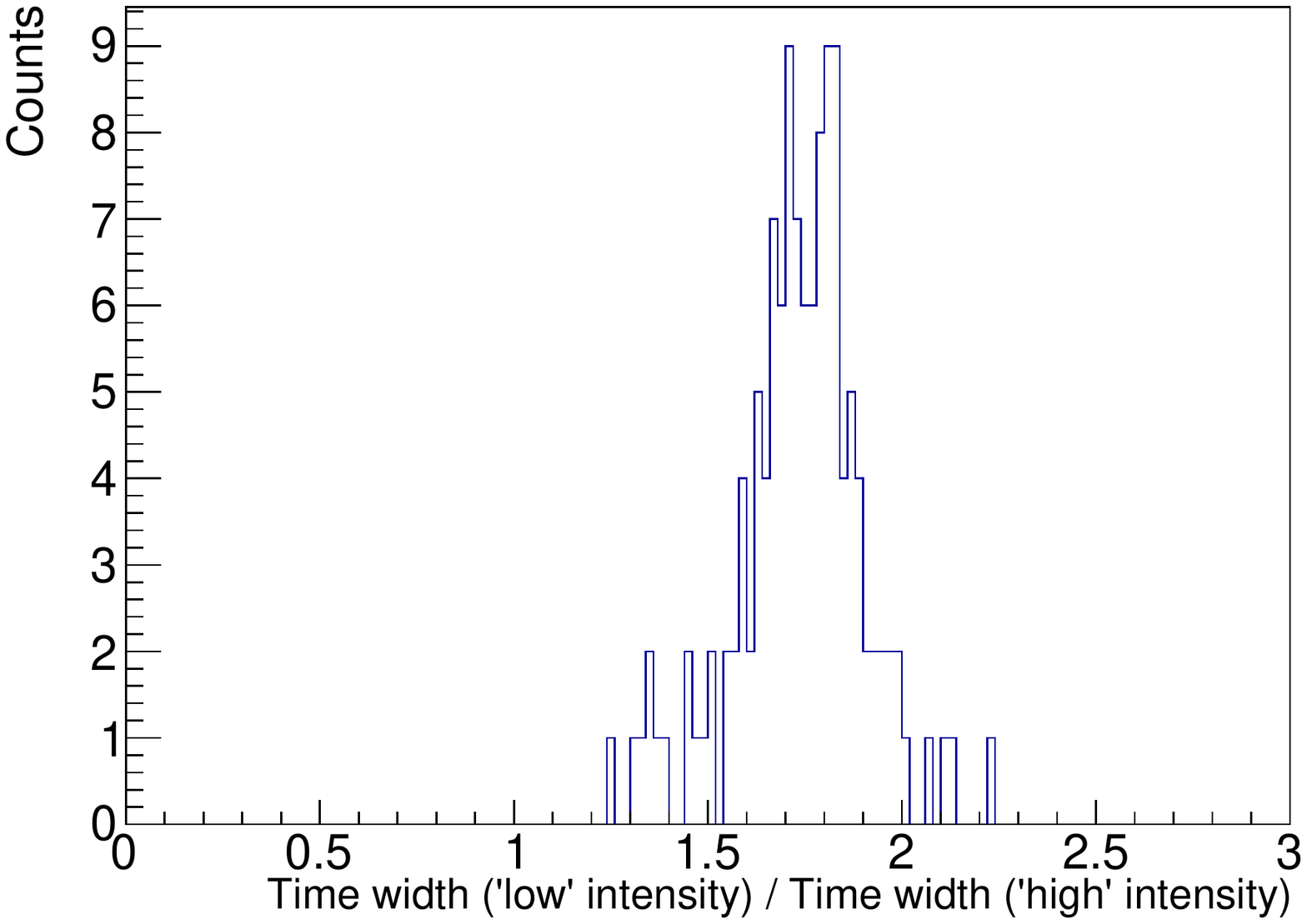} \par %
        \includegraphics[width=1.0\linewidth]{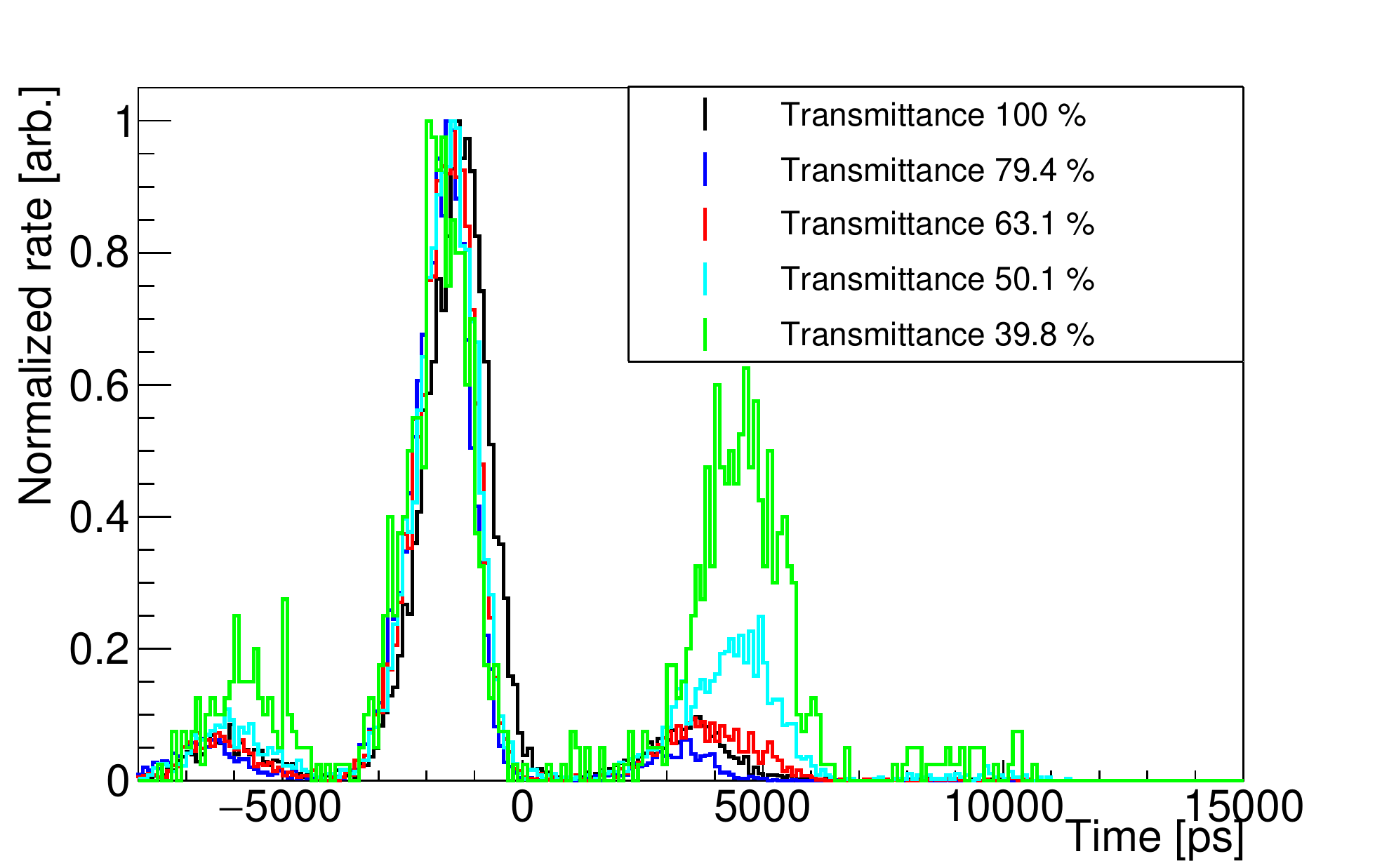} \par 
\end{multicols}
\vspace{-0.5cm}
\caption{Timing measurements:  
(a) time spread of only one pixel [SiPM A (X,Y) : (3,1)] with respect to the reference pixel [SiPM B (X,Y) : (3,3)] with laser intensity varied across all pixels at the same time; 
(b)~ratio of time widths for 127 pairs of signal with respect to the reference pixel  of between dim and bright signals (marked as 'low' and 'high' intensities, shown as blue and purple lines on the subfigure (a) for one pair of the pixels); (c)~Timing between two pixels (one pixel had a varying ND filter, as shown in Fig.~\ref{fig_laser_setup}. In the  subfigures (a) and (c), the $\sim$5-ns artifacts are due to the electronics readout internal 200-MHz clock.}
\label{fig_timing_tests}
\end{figure*}

\subsection{Characterization of detector timing}
Neutron double-scatter imaging requires accurate knowledge of the neutron energy after the first scatter, which can be obtained from a time-of-flight measurement between two neutron scatters. 
For time of flight, we need to ensure a relative timing calibration for all of the channels, since each channel can have different timing offsets and/or resolutions.

To measure timing offsets, we directed a  collimated fan-like beam of gamma rays at the center of the scintillator array so that the relative timing of the two pixels at either end of each scintillator rod can be measured. An example of two pairs of pixels (viewing two rods) is shown in Fig.~\ref{fig_timing_offsets_2rods_laser_Cs}.
For the example shown, we selected two rods (two pairs of pixels) with a large difference in offset.

In practice, it is difficult to compare the timing of pixels unless they are paired by a single scintillator rod.
Therefore, this method cannot be used to obtain the timing offsets of all the possible pairs of pixels. 
For this purpose, we developed a laser setup that can illuminate all the pixels of each SiPM array simultaneously, as shown in Fig.~\ref{fig_laser_setup}. 
The laser 
(Picoquant DL-800D with a 405-nm laser head LDH-P-C) has a pulse rise time of $\leq$25~ps. 
However, the light is propagated via a 1-meter multimode fiber from the laser head to the light-tight enclosure, which has the potential to smear the rise time. 
To illuminate the pixels uniformly, the light was directed from the end of the fiber through a collimator and then a diffuser consisting of several layers of PTFE tape.

It was measured that the standard deviations in the timing difference between Total and Head channels, viewing the same pixel, are $\leq$100 picoseconds, which means that it is possible to use either Head or Total timestamps in the analysis; for clarity, we only use timing obtained with Total channels.

The timing uncertainty between any SiPM pixel and a reference pixel (pixel 3/3 on SiPM array B) was found to be between $\sim$500--1000~ps (Fig.~\ref{fig_T_width_map_A}). 
One important test is to ensure that the relative timing offsets as measured by the laser are the same as those measured by the center-collimated gamma-ray source, as shown in Fig.~\ref{fig_timing_offsets_2rods_laser_Cs}.
It can also be seen that the width of the distributions is dependent on the brightness of the signal. %
First, we show an example of the timing dependence on  intensity. In Fig.~\ref{fig_timing_tests} the time difference between two pixels is plotted for six different laser intensities, while the values of mean charges of those pixels are reported in the legend. 
For brighter laser settings the timing distributions get narrower; this trend holds for the majority of pixels. Figure~\ref{fig_timing_tests}b shows the distribution of ratios for  timing widths for 127 pixels relative to a single reference pixel using two different laser brightness settings.

We also used a set of neutral-density filters to explore a scenario where the relative light intensity between a pixel and the reference pixel changes; this is shown in Fig.~\ref{fig_laser_setup}.
As the brightness observed by the pixel changes, the offsets move by a few hundred picoseconds (Fig.~\ref{fig_timing_tests}c).

\begin{figure}%
  \centering
  \includegraphics[width=.5\textwidth]{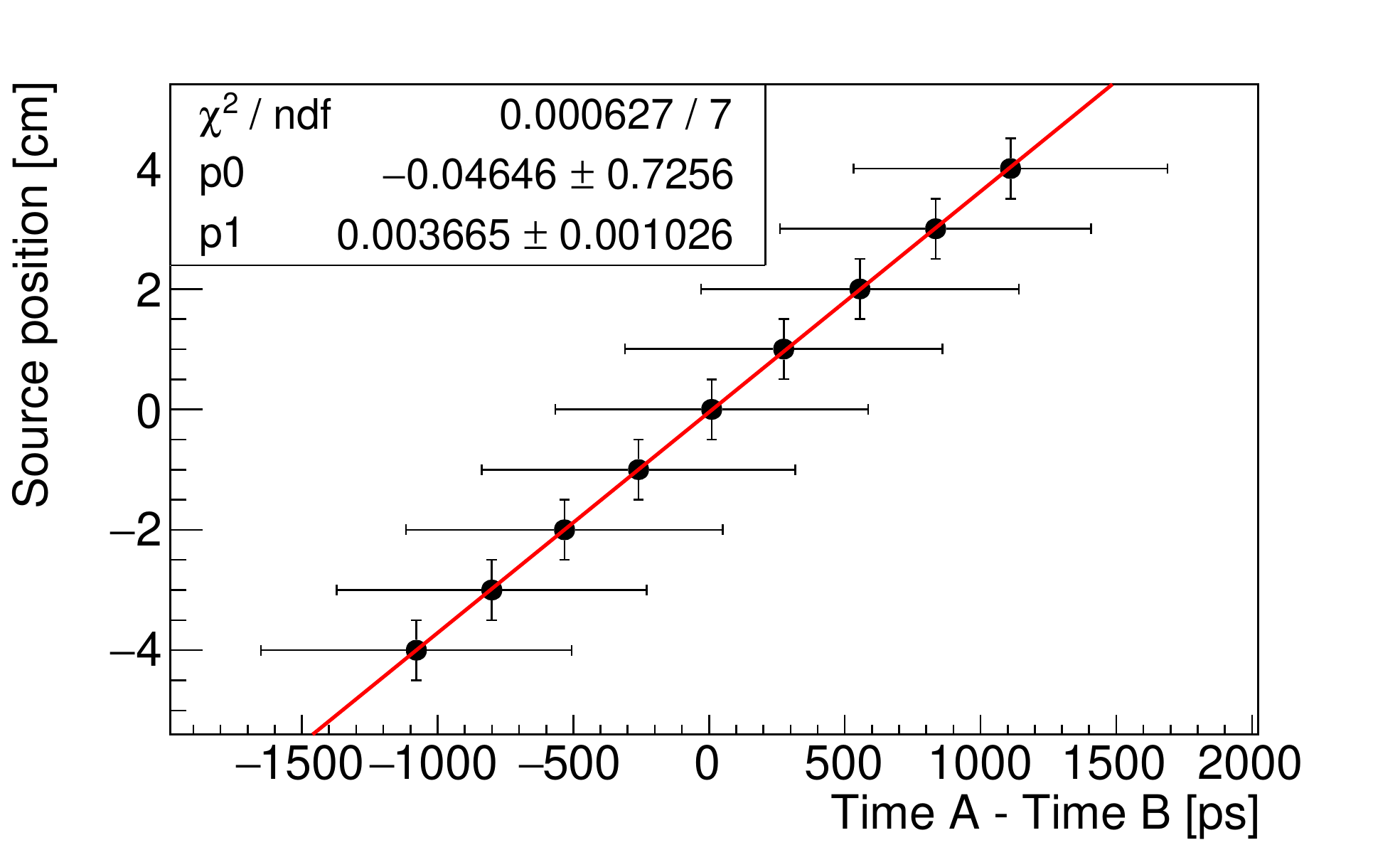} \par 
  \caption{Position reconstruction along the rods using timing difference between pixels on the opposite sides using a collimated $^{137}$Cs source}
  \label{fig_position_vs_timing}
\end{figure}

After we calibrated the detector, it was possible to reconstruct position using timing as shown in Fig.~\ref{fig_position_vs_timing} and using $^{137}$Cs data.

\subsection{Position resolution}

Given the previous measurements of timing and energy resolution, it is worth examining the position uncertainty for the iSANDD along the $z$-axis. 
The scintillator used in this work is $^6$Li-doped PSD plastic with a light yield of approximately 9,000 photons per MeV~\cite{Sutanto:2021xpo}. For the iSANDD, the scintillator bars were about the same length of the first SANDD prototype~\cite{Li:2019sof}, which used undoped PSD plastic. The yield of all three detector configurations were consistent with commercial plastic (within 10\%).

\begin{figure}[ht!]
    \includegraphics[width=1.\linewidth]{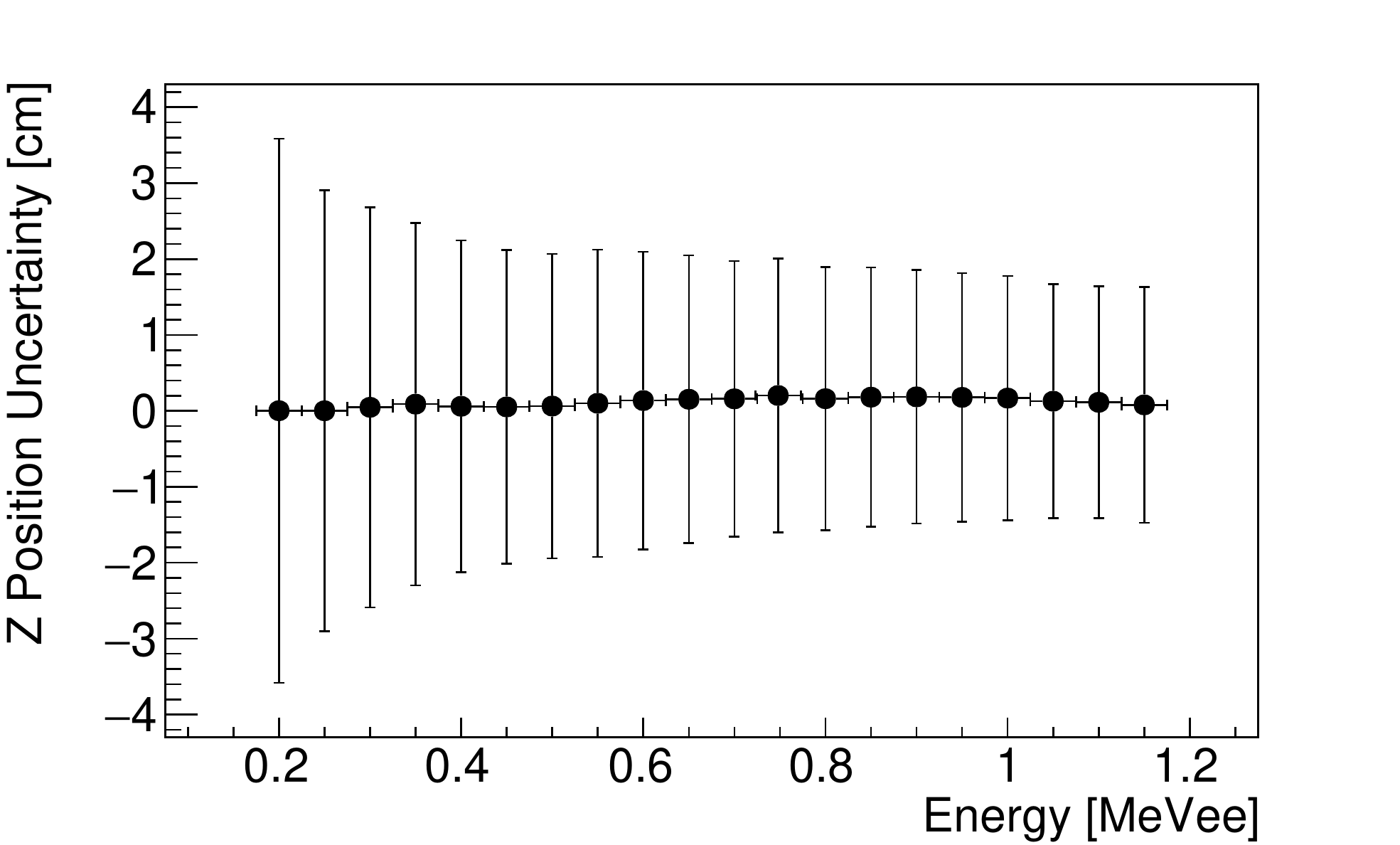}
    \vspace{-\mylength}
    \begin{center}
        Time-based:
    \end{center}
\vspace{\mydelt}
    \includegraphics[width=1.\linewidth]{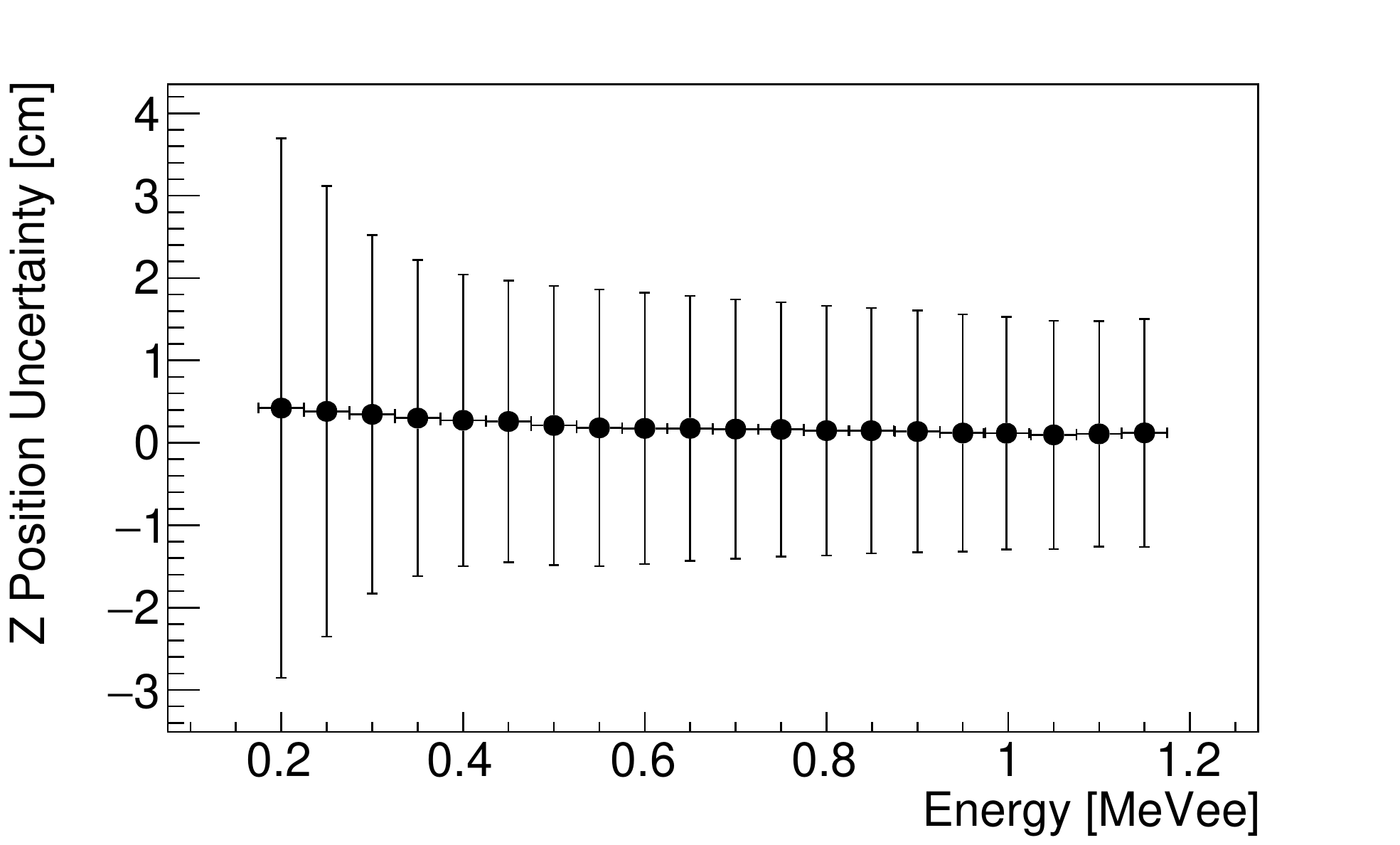}
    \vspace{-\mylength}
    \begin{center}
        Charge-based:
    \end{center}
\vspace{\mydelt}
\vspace{-0.5cm}
\caption{The $z$-position uncertainty calculated as a function of energy --- time-based (top panel) and charge-based (bottom panel), both with using the $^{22}$Na data. The charge-based uncertainty is slightly smaller compared to the time-based one.}
\label{fig_gamma_Zuncertainty_vs_E}
\end{figure}

We examined the $z$-position resolution that results from timing and energy separately.
For energy, we can reconstruct $z$-coordinate of the interaction using charge  ratio $R_{AB}$, as shown in Fig.~\ref{fig_diffZ}.
From the figure, the estimated position resolution measured with a $^{137}$Cs source was $\sim$2 cm.
The position uncertainty as a function of energy measured with a collimated $^{22}$Na source is shown in Fig.~\ref{fig_gamma_Zuncertainty_vs_E}.
When comparing these results with the earlier SANDD prototype~\cite{Li:2019sof}, which used full-waveform digitization, the position resolution observed here was $\sim$50\% worse over the energy range of 0.2--1.2~MeVee.

The position resolution due to timing was also measured.
The position resolution using the $^{22}$Na  data is shown in Fig.~\ref{fig_gamma_Zuncertainty_vs_E}.

\deflen{mylength}{175pt}
\deflen{mydelt}{145pt}

\subsection{Pulse-shape discrimination}
As mentioned previously, a pulse-shape parameter (PSP) was defined using the Total and Head charges from each pixel:
\begin{equation}
    PSP = \frac{Q_{tail}}{Q_{total}} 
    = \frac{Q_{total} - Q_{head}}{Q_{total}}.
\end{equation}
Using a $^{252}$Cf source, the iSANDD proved insensitive to particle-interaction type via PSD, unlike earlier measurements taken with  full-waveform digitizer readout~\cite{Li:2019sof}. Since the characteristic of the scintillator and the photodetectors were similar in both cases, the significant deterioration in PSD sensitivity is attributed to the electronics readout.

\begin{figure}%
  \centering
  \includegraphics[width=1.0\linewidth]{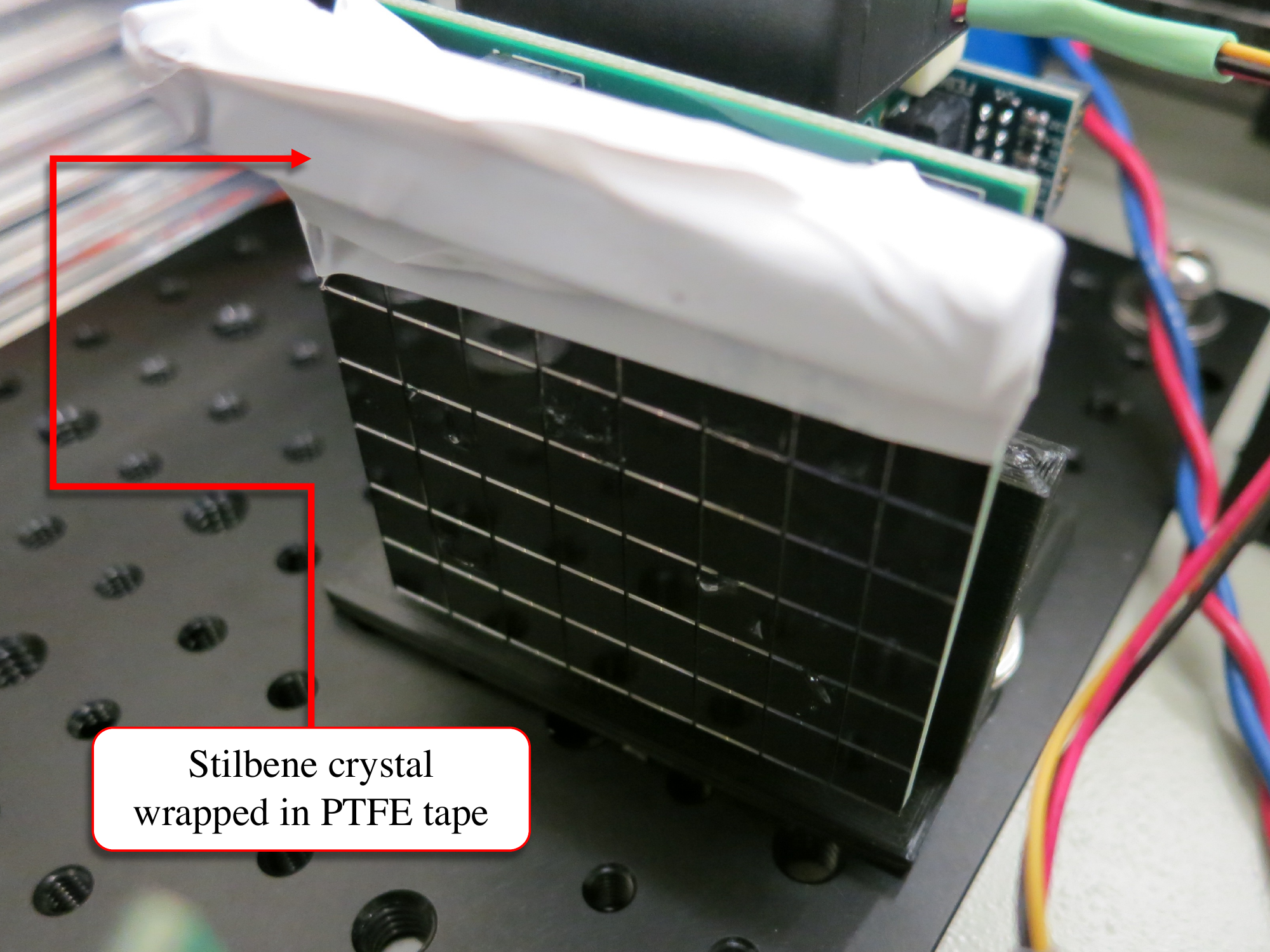}
  \caption{A stilbene crystal (6~mm $\times$ 6~mm $\times$ 50~mm) wrapped in PTFE tape on five sides and optically coupled to 8 pixels of the SiPM array via EJ-550 silicon grease.}
  \label{fig_stilbene_setup}
\end{figure}

To further investigate PSD sensitivity with the PETsys readout, a stilbene crystal was coupled to the SiPM array, as shown in Fig.~\ref{fig_stilbene_setup}. 
This orientation and the stilbene crystal as the scintilator were chosen wiht a goal to improve the PSD performance. 
 
\begin{figure*}[ht!]
\begin{multicols}{3}
\begin{centering}
    Stilbene ($^{252}$Cf) \par 
    Stilbene ($^{137}$Cs)\par 
    Plastic ($^{252}$Cf) \par 
\end{centering}
\end{multicols}
\vspace{-.8cm}
\begin{multicols}{3}
    \includegraphics[width=1.0\linewidth]{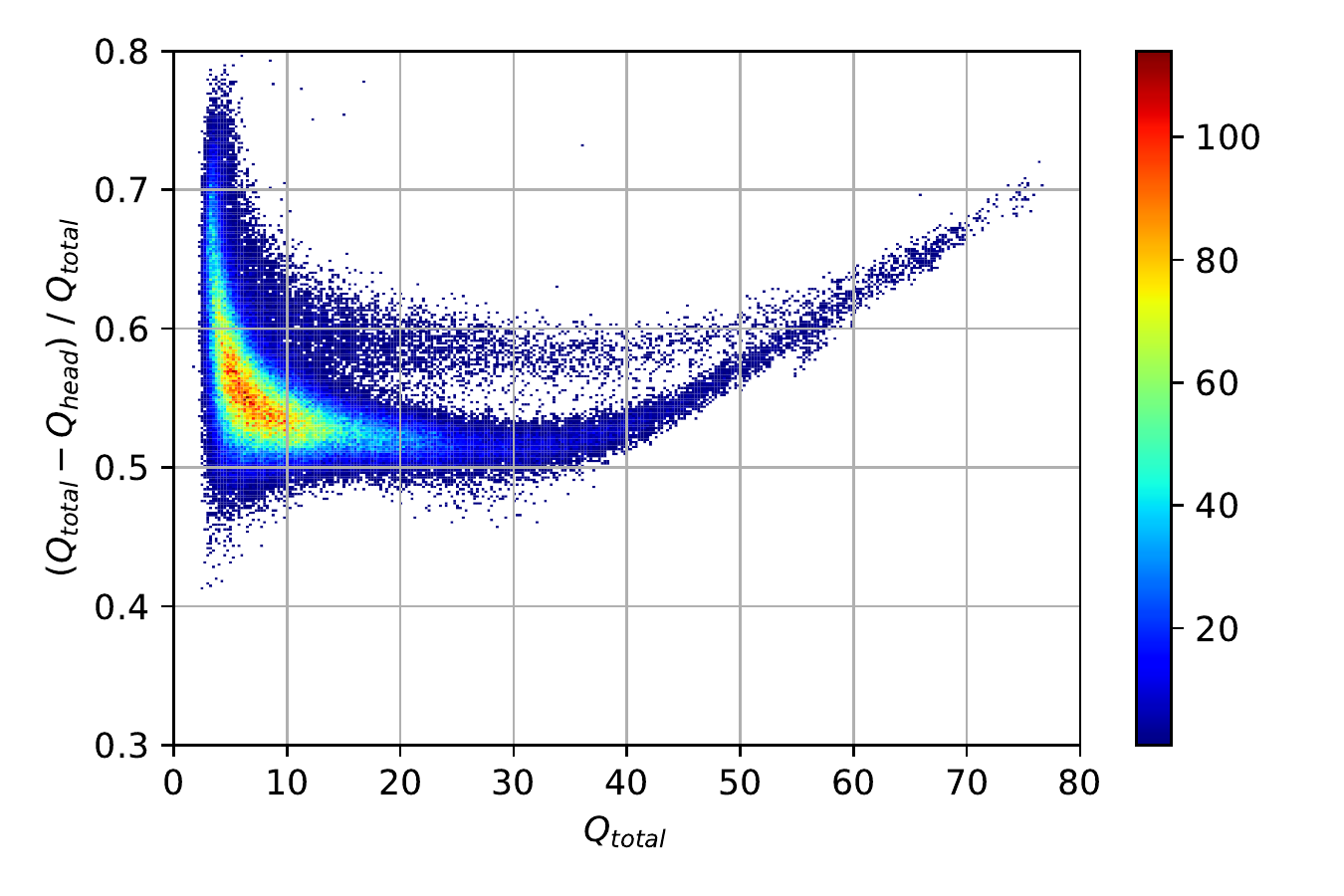} \par 
    \includegraphics[width=1.0\linewidth]{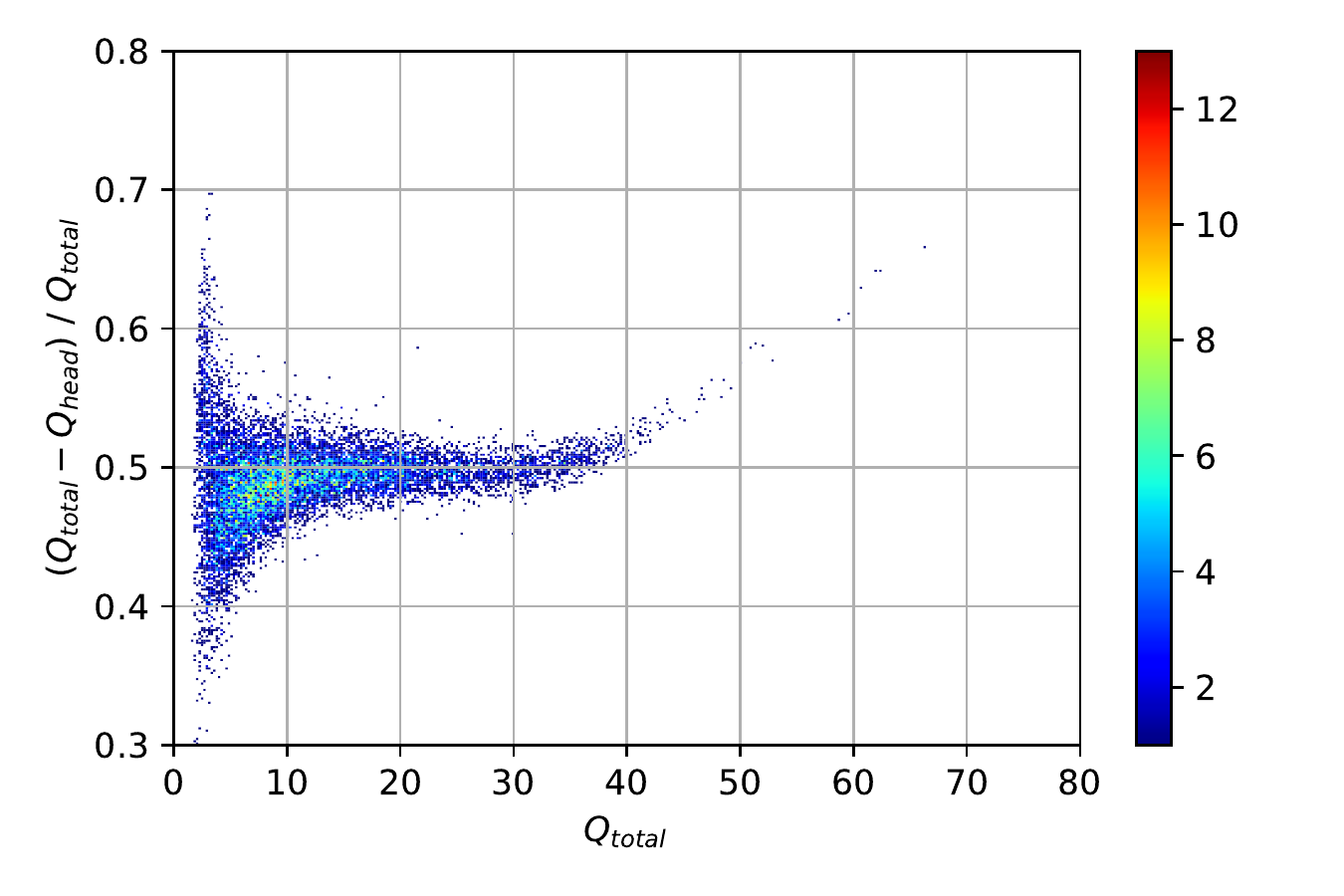} \par 
    \includegraphics[width=1.0\linewidth]{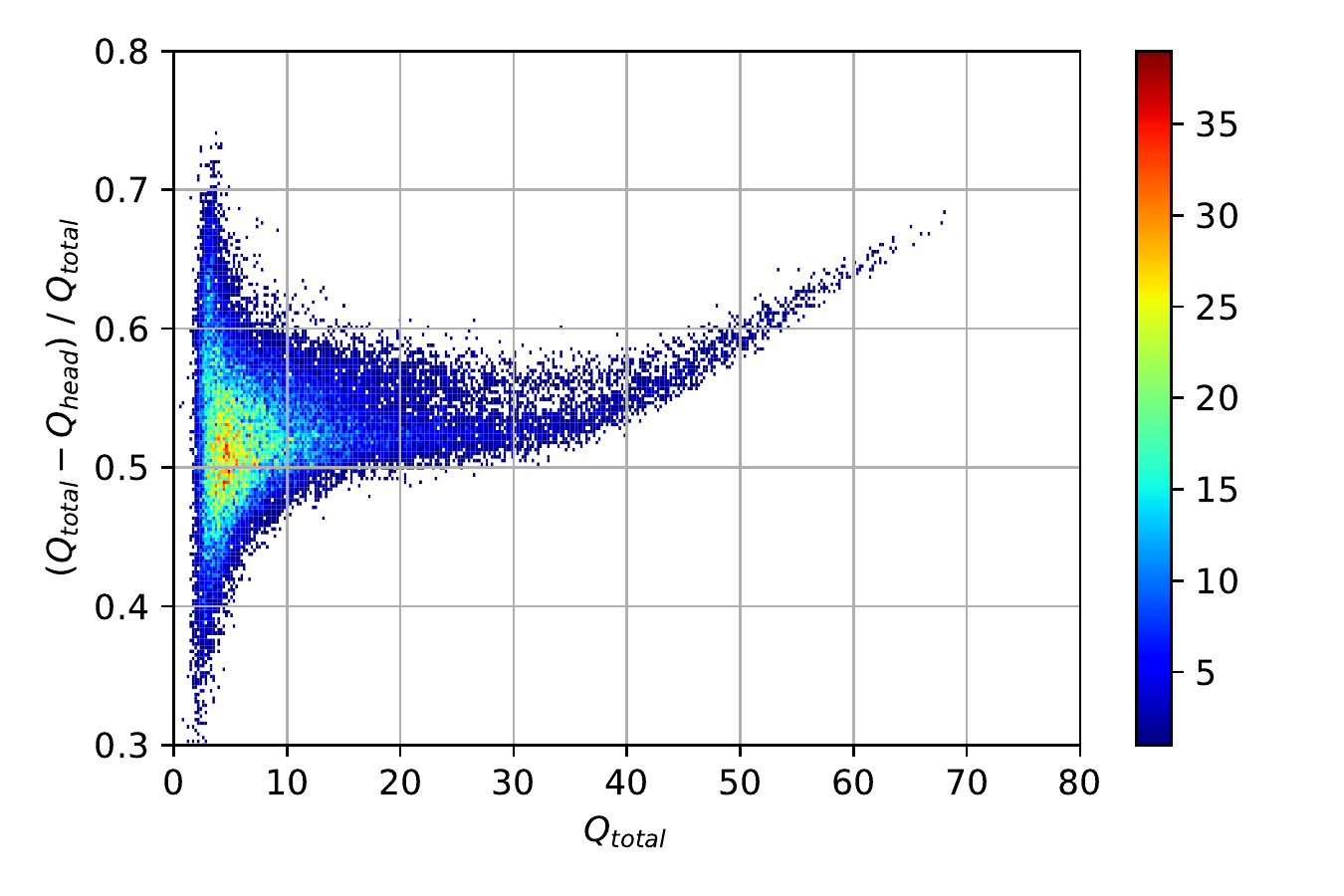} \par 
\end{multicols}
\vspace{-0.5cm}
\caption{Examples of detector response for stilbene and PSD plastic, for the same pixel. Both the stilbene crystal and the plastic rod were wrapped in teflon and oriented as shown in Fig.~\ref{fig_stilbene_setup}. Units of charge are PETsys ADC units. The gamma-ray band in the stilbene $^{252}$Cf plot has a slightly higher PSD value, which might be due to a $\sim$10 times higher event rate.}
\label{fig_PSD}
\end{figure*}

 After switching to stilbene, modest improvement in PSD sensitivity was observed, as shown in Fig.~\ref{fig_PSD}. However, swapping the stilbene crystal for the PSD plastic scintillator in the same orientation showed only modest PSD sensitivity.

One problem that may be impacting sensitivity to PSD is that, at present time there is a lack of flexibility in terms of the range of integration times for Head and Total.
For example, the Head integration time cannot be set below $\sim$100~ns, and the Total integration time cannot be set greater than $\sim$400~ns. 
It is hoped that future iterations of the PETsys ASIC and the splitter board would allow for greater flexibility in integration time selection.

\begin{figure*}[ht!]
\begin{multicols}{3}
\begin{centering}
    (10 neutrons) \par 
    (100 neutrons) \par 
    (829 neutrons) \par 
\end{centering}
\end{multicols}
\vspace{-.8cm}
\begin{multicols}{3}
    \includegraphics[width=1.0\linewidth]{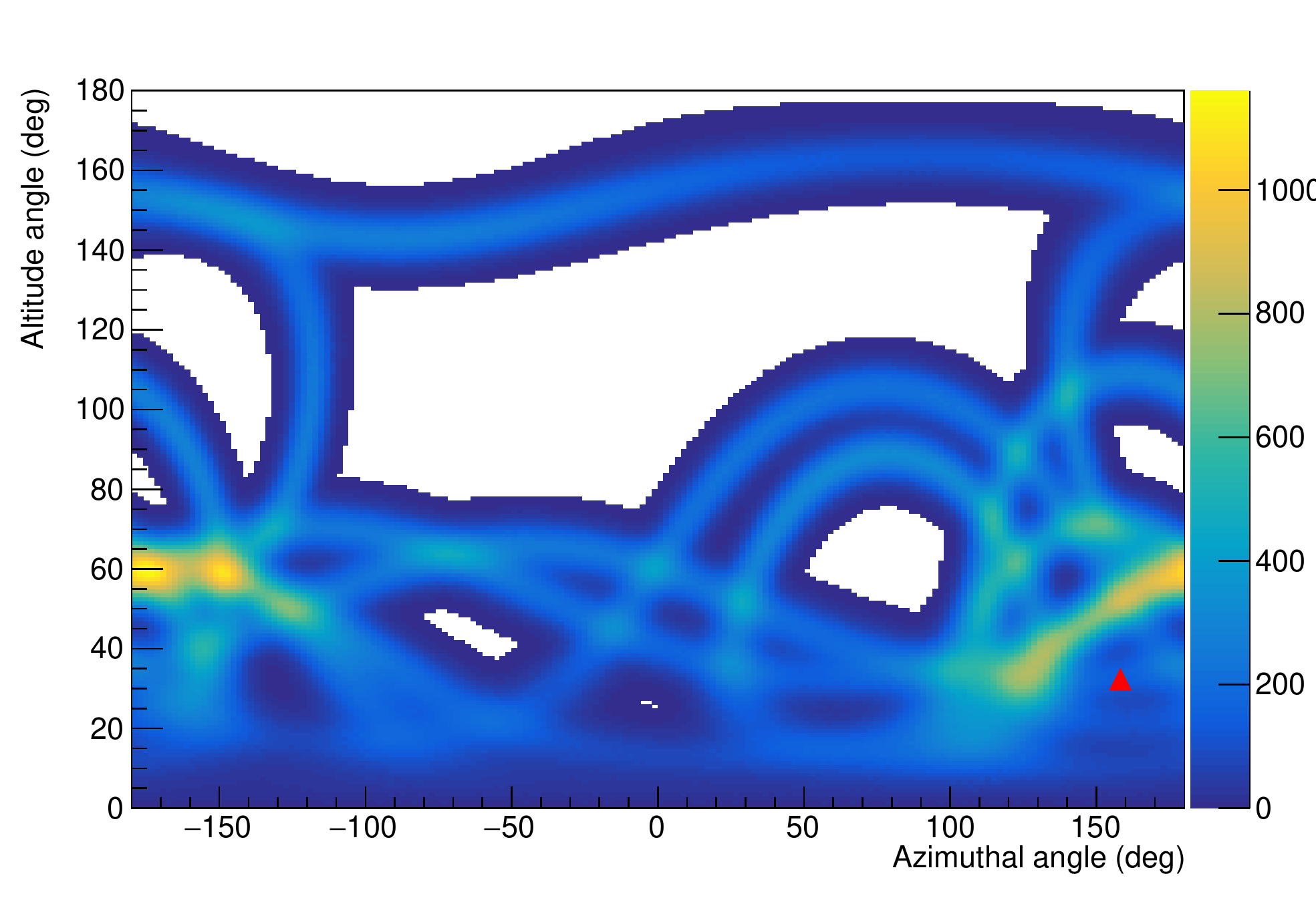} \par 
    \includegraphics[width=1.0\linewidth]{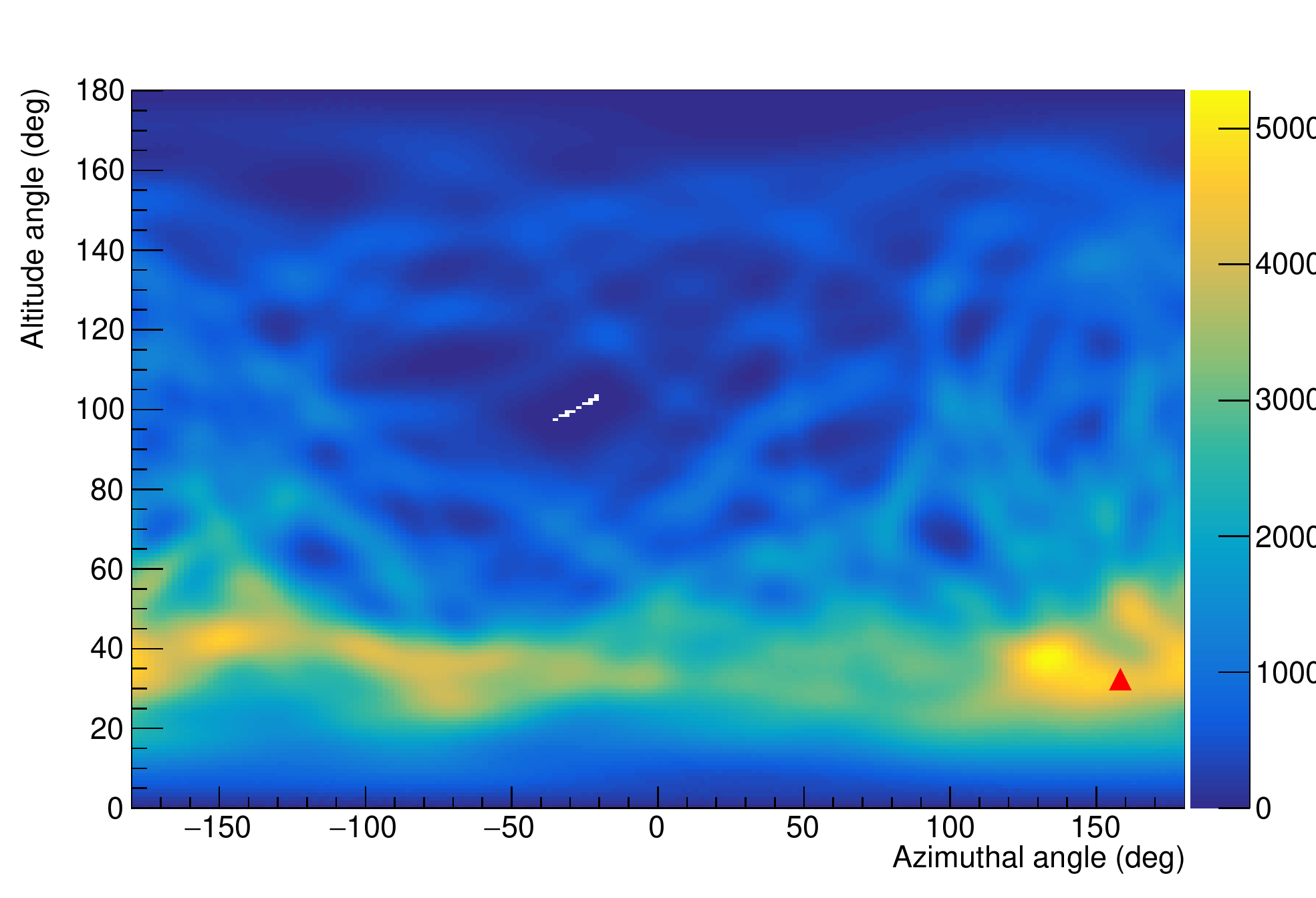} \par 
    \includegraphics[width=1.0\linewidth]{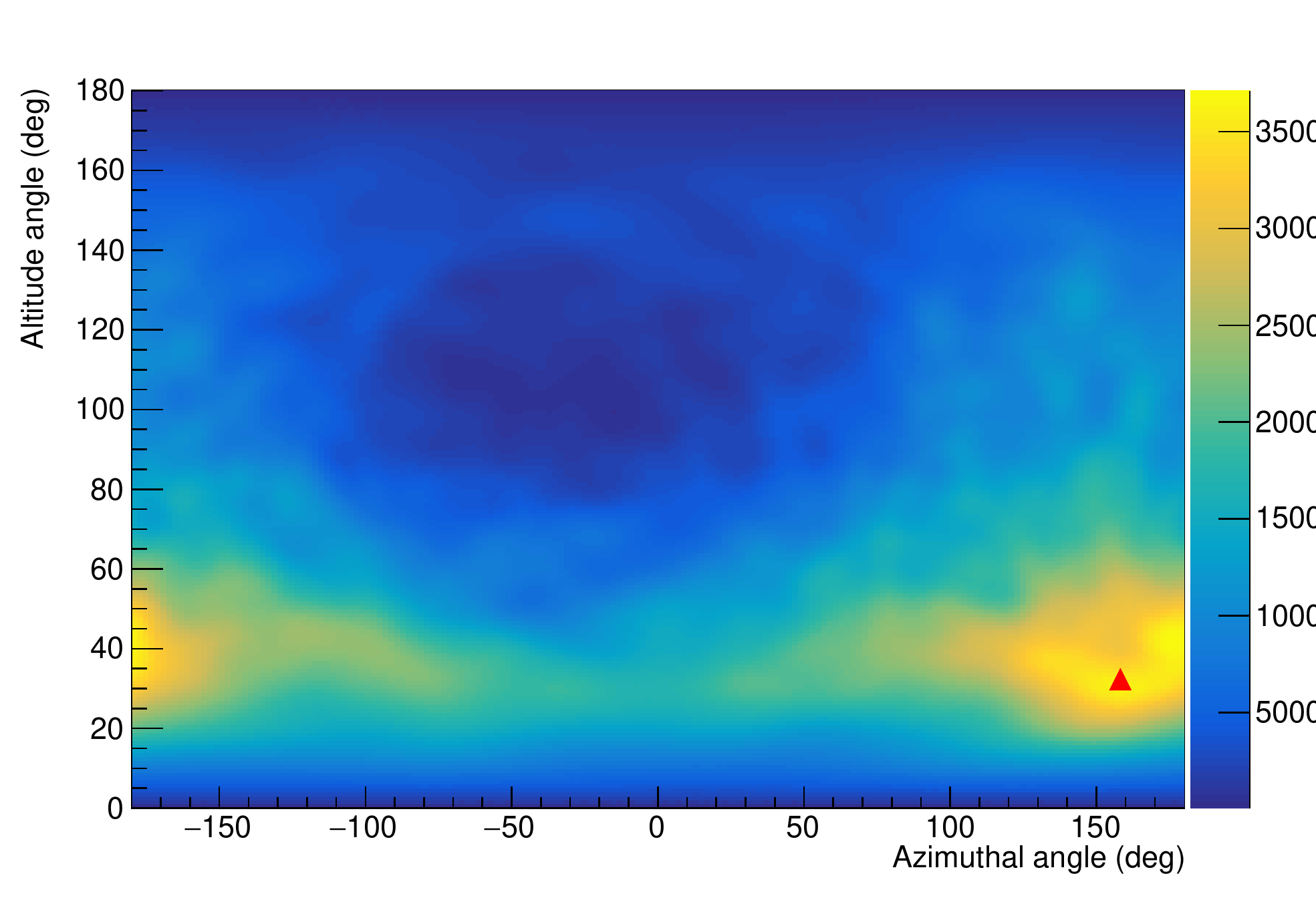} \par 
\end{multicols}
\vspace{-0.5cm}
\caption{Neutron directionality maps based on a simulation of 2.45-MeV neutrons scattering twice within the detector. The red triangle indicates the true position of the source (azimuth = 158.2$^\circ$, altitude = 32.4$^\circ$). Each double-scatter event represents a ring on the $4\pi$ sky map. As the number of double scatters increases (from left to right), the true position of the source is more clearly revealed particularly as the number of double-scatters increases from 10 to 100. Gaussian smearing is applied. The selection criteria are described in the text.}
\label{fig_neutron_directionality}
\end{figure*}

\subsection{Assessing neutron-imaging capabilities}
\label{sec_simulation}

In this section, we predict the imaging capabilities of the PETsys system using the measured energy, timing, and position resolution, while assuming that some degree of PSD sensitivity improvement can be realized.
This analysis can help determine whether the energy, timing, and position resolutions are sufficient for practical neutron imaging.

A Geant4-based~\cite{geant4}
simulation was performed, where the detector performance parameters measured with the PETsys readout were used as input parameters.
We also assumed a quenching factor, such that a proton recoil of 1~MeV is seen as $\sim$0.16~MeVee, while a 2-MeV recoil --- $\sim$0.5~MeVee. %
Figure~\ref{fig_neutron_directionality} shows 
the result of this simulation for monoenergetic 2.45-MeV  neutrons for a random selection of varying numbers of double-scattered neutrons. 
The cone with an opening angle $\theta$ of possible incoming neutron direction is defined as
\begin{equation}
    \tan \theta =  \sqrt{\frac{ E_p }{ E_{\mathrm{TOF}} } }
    = 
    \frac{t_2 - t_1}{\left| \mathbf{r}_2 - \mathbf{r}_1 \right|} \sqrt{\frac{ 2  E_p }{ m } },
\end{equation}
where $m$ is the neutron mass, $E_p$ is the proton kinetic energy, and $(\mathbf{r}_1, t_1)$ and $(\mathbf{r}_2, t_2)$ are positions and times of the 1$^\mathrm{st}$ and 2$^\mathrm{nd}$ scatters. 
The timing, energy, and position uncertainties are propagated to calculate the width of each neutron-scatter ``ring''.

The following cuts were set:
segment multiplicity = 2; 
time difference between segment hits $>$1~ns; 
quenched energy deposit in each segment $>$0.3~MeVee; 
distance between segment hits $>$3~cm.
Energy is quenched based on a LS quenching profile and smeared for photostatistics based on 20\%$/\sqrt{E}$.
Hit times are smeared by 750~ps;
Hit positions are given by the $x$ and $y$ centers of the segment hit, and with a 2~cm smearing on the position along the rod ($z$-coordinate). 
The simulation generated $5 \times 10^6$ 2.45-MeV neutrons in a cone aimed at the detector from one meter away such that the opening angle had $\cos\theta=0.99$.  This corresponds to a $\sim$10$^9$ isotropic neutron source at a 1~m distance.

\section{Conclusion}

We assessed the suitability of one example of an ASIC-based miniaturized data acquisition electronics readout, developed for PET-imaging, for neutron imaging applications. 
The readout was tested with a 64-rod $^6$Li-doped PSD plastic scintillator prototype, coupled to two 64-pixel SiPM arrays.  The 128 individual SiPM pixels were each split into a short- and long-integration window, and read out by 256 independent electronics channels. The short- and long-integration windows were combined to create a PSD metric. 
Energy and timing measurements indicated resolutions of $\sim$20\% (at the $^{137}$Cs Compton edge) and $\sim$750~ps, respectively. 
The $z$-position resolution was measured to be $\sim$2~cm using either timing or the charge ratio $R_{AB}$. 
The position resolution for $(x,y,z)$ was measured as $(0.5, 0.5, 2)$~cm.
The resolutions in $x$ and $y$  are 0.5~cm due to segmentation. 

The $z$-position uncertainty using the charge ratio ($R_{AB}$) was somewhat worse than  our previous measurements of a similar prototype with full-waveform digitization~\cite{Li:2019sof}.
This reduction in position sensitivity is likely caused by the worsening in energy resolution. 
We theorize that this may be partly due to the splitting of the SiPM signals into two independent electronics channels used to integrate Head and Total charge.

The $z$-position resolution based on timing was a factor $\sim$1.4 worse than recently reported measurements using fast full-waveform digitizers and fast commercial plastic scintillators~\cite{Keefe:2021vhk}.
Since the plastic scintillator used in this work is not optimized for speed, it is possible that the scintillator contributes somewhat to the degraded position resolution.
The worsening due to the electronics readout alone can therefore be constrained to $\lesssim$1.4.

To achieve PSD performance sufficient for reliable gamma-ray background rejection, further improvements to the electronics readout scheme will be needed. 
One possible cause might be the increased noise associated with the Head and Total charge-splitting mentioned earlier. 
It is possible that the splitting of the signal from each pixel is a contributing factor to noise and therefore inferior PSD.
Another possible avenue for improvement is waveform digitization by ASIC-based readouts. 

Another problem we have identified with the PETsys readout is a tendency for the baseline (the pedestal) to shift as a result of temperature fluctuations.
This problem can impact the apparent energy resolution if the temperature shifts over the course of a data acquisition run.
We did not investigate the impact of temperature fluctuations on PSD.
In a field application of iSANDD, one would have to implement an algorithm that determines the baseline and gain for every channel as a function of temperature. A suitable way to monitor baseline shifts would be to trigger the electronics channels externally in order to set a pedestal. 

To estimate the likely angular sensitivity of  a neutron-scatter camera with energy, timing, and position resolutions as presented here, we performed simulations of a flux of DD-like fast neutrons incident on the camera. 
These simulations indicated adequate directional sensitivity from 100 neutron double-scatter events. Only limited improvements in angular sensitivity were gained from  $\gtrsim$100 such events.

\section{Acknowledgements}
The scintillator used in this project was synthesized by \mbox{Andrew~Mabe}, and is of the same composition as used in the SANDD~\cite{Sutanto:2021xpo}.
We thank Natalia Zaitseva, Glenn Jocher, and Melinda Sweaney for useful discussions, Nathaniel Bowden for the interest in the project, and Dom Porcincula for designing and 3D-printing various mechanical parts used in this project. We further acknowledge the discussions with PETsys support staff and with Shiva Abbaszadeh. 

This work was supported by the U.S. Department of Energy National Nuclear Security Administration and Lawrence Livermore National Laboratory [Contract No. DE-AC52-07NA27344, LDRD tracking number 21-FS-025, release number LLNL-JRNL-830322] and the Consortium for Monitoring, Technology, and Verification (DE-NA0003920). The research of T. C. W. was partially supported by the Department of Energy, Nuclear Energy University Program Fellowship.

\bibliographystyle{elsarticle-num}

\bibliography{ref}

\end{document}